\documentclass[useAMS,usenatbib]{mn2e}
\usepackage[dvipdfm]{graphicx}
\usepackage{times}
\usepackage{amssymb}
\usepackage{amsmath}
\usepackage{rotating}
\usepackage{multirow}
\usepackage{epsfig}
\input{epsf}

\title[The origin of ultra-fast outflows in AGN: Monte-Carlo simulations
of the wind in PDS~456]
{The origin of ultra-fast outflows in AGN: Monte-Carlo simulations
of the wind in PDS~456}

\author[K. Hagino et al.]
{Kouichi Hagino$^{1,2}$, Hirokazu Odaka$^{1,3}$, Chris Done$^4$, Poshak Gandhi$^4$, Shin Watanabe$^{1,2}$,
\newauthor
Masao Sako$^5$ and Tadayuki Takahashi$^{1,2}$\\
$^1$ Institute of Space and Astronautical Science (ISAS), Japan Aerospace Exploration Agency (JAXA), 3-1-1 Yoshinodai, Chuo, Sagamihara, Kanagawa 252-5210, Japan\\ 
$^2$ Department of Physics, University of Tokyo, 7-3-1 Hongo, Bunkyo, Tokyo 113-0033, Japan\\
$^3$ Max-Planck-Institut f\"{u}r Kernphysik, P.O. Box 103980, D 69029 Heidelberg, Germany\\
$^4$ Department of Physics, University of Durham, South Road, Durham DH1 3LE, UK\\
$^5$ Department of Physics and Astronomy, University of Pennsylvania, 209 South 33rd Street, Philadelphia, PA 19104, USA
}

\date{Submitted to MNRAS}
\pagerange{\pageref{firstpage}--\pageref{lastpage}} \pubyear{}

\begin{document}

\topmargin = -0.5cm

\maketitle

\label{firstpage}

\begin{abstract}
Ultra-fast outflows (UFOs) are seen in many AGN, giving a possible mode for
AGN feedback onto the host galaxy. However, the mechanism(s) for the
launch and acceleration of these outflows are currently unknown, with
UV line driving apparently strongly disfavoured as the material along
the line of sight is so highly ionised that it has no UV transitions.
We revisit this issue using the {\it Suzaku} X-ray data from PDS~456,
an AGN with the most powerful UFO seen in the local Universe.  We
explore conditions in the wind by developing a new 3-D Monte-Carlo
code for radiation transport. 
The code only handles highly ionised ions, but the data show the ionisation state of the wind is high enough that this is appropriate, and this restriction makes it fast enough to explore parameter space.
We reproduce the results of earlier work, confirming that the mass loss rate in the
wind is around 30\% of the inferred inflow rate through the outer
disc. 
We show for the first time that UV line driving is likely to be a major contribution to the wind acceleration. The
mass loss rate in the wind matches that predicted from a purely line driven system, and this UV absorption can
take place out of the line of sight. Continuum driving should also play a role as the source is close to Eddington.
This predicts that the most extreme outflows will be produced from the highest mass accretion rate flows onto high mass black holes, as observed. 
\end{abstract}


\section{Introduction} \label{sec:introduction}

AGN-driven winds are potentially the most effective way of
transporting energy and momentum from the nuclear scales to the host
galaxy, quenching star formation in the bulge by sweeping away the gas
reservoir. This feedback process can quantitatively reproduce the
$M-\sigma$ relation (e.g. \citealt{King2010}).

We see clear observational evidence of winds in AGN via absorption
lines. In the UV and X-ray bands we observe narrow absorption lines
outflowing with moderate velocity of hundreds to few thousand
km~s$^{-1}$. This warm absorber is detected in ~50\% of AGN
\citep{Blustin2005,Piconcelli2005,McKernan2007}, and
may have its origin in a swept-up ISM or thermally driven wind from
the molecular torus (\cite{Blustin2005}). However, this carries only a
small fraction of the kinetic energy, as the amount of material and
outflow velocity are both quite small (e.g \citealt{Blustin2005}). 

Instead, there are two much higher velocity systems which potentially
have much greater impact on the host galaxy. In the UV band, broad
absorption lines (BAL) are seen in $\sim30$\% of AGN, and may be present but
outside the line of sight in most AGN \citep{Ganguly2008,Elvis2000}.
These absorbers can be outflowing as fast as $\sim 0.2c$, so carry
considerable kinetic energy, and probably arise in a UV line driven
wind from the accretion disc (e.g. \citealt{Proga2004}).

However, the most powerful outflows appear to be so highly ionised
that the only bound transitions left are for Hydrogen- and Helium-like
iron.  Such winds can only be detected at X-ray energies, and a few
AGN have substantial columns of material outflowing at speeds of up to
$\sim 0.3c$ \citep{Tombesi2010,Gofford2013}, and in a handful of
higher redshift AGN at up to 0.7c \citep{Chartas2002,Lanzuisi2012}.
These high velocities point to an origin very close to
the SMBH, but the launching and acceleration mechanism remain unclear.
Possibilities include radiation-driven winds as the source
approaches/exceeds Eddington \citep{King2010} and/or magnetic driving
(e.g. \citealt{Blandford1982}), but UV line driving is generally 
not thought to be important as 
the high ionisation state of the material means it has
negligible UV opacity \citep{Tombesi2013}.

The lack of insight into the wind acceleration mechanism means that
even the best wind models are somewhat ad-hoc, and impose a geometry
and velocity structure on the wind. The wind is probably not
spherical \citep{Elvis2000}, so the radiative transfer cannot be modelled 
analytically via the Sobolov approximation. Instead, the best current codes
do full Monte-Carlo radiative transfer through the wind material,
solving also for the ionisation balance at each point in the wind
\cite{Sim2008,Sim2010a,Sim2010b}. However, such a detailed ionisation
calculation is slow, so exploring parameter space is difficult. 

Here we develop a new Monte-Carlo code, using only the H- and He-like
ion stages (see also \citealt{Sim2008,Sim2010a}) so that it is fast.
We use this to fit to PDS~456 ($z=0.184$), one of the most luminous
objects in the local Universe ($z<0.3$). This is intrinsically of
similar luminosity in the optical than 3C~273, though it is heavily
absorbed by E(B-V)=0.48 as it lies close to the plane of our Galaxy
\citep{Simpson1999}. This also hosts the most powerful outflow known
in the local Universe
\citep{Reeves2003,Reeves2009,Tombesi2010,Gofford2013}, lending support
to the radiation driven wind models since the luminosity is close to
Eddington for its $\sim 2\times10^9$~$\mathrm{M_\odot}$ black hole
\citep{Reeves2009}.

We use {\it Suzaku} data of PDS~456 \citep{Reeves2009,Reeves2014,Gofford2014} for this work
because it has a low and stable background and the best spectral resolution with a relatively large collecting area in the Fe K band.
Thanks to these capabilities, {\it Suzaku} is best suited to study the highly blue-shifted Fe K absorption lines.

The wind in PDS~456 has previously been studied using the \cite{Sim2010a} code by \cite{Reeves2014}. We obtain similar results
for similar parameters, demonstrating that the new code is reliable,
but we are also able to use our fast code to explore a wide range of
parameter space, and show how the observed properties of the wind
change with each physical parameter.
We reproduce all the {\it Suzaku} observations of PDS~456, and we can explain the time variability of the wind spectrum in the single modelling framework. 
We speculate that the wind is launched by a combination of UV
line driving and radiation pressure, but that the UV line driving region is close to the disc, out of the line of sight. 

Below, we assume a standard cosmology with $H_0=71$\,km\,s$^{-1}$\,Mpc$^{-1}$, $\Omega_{\rm m}=0.27$ and $\Omega_\Lambda=0.73$, so that the redshift of the target $z = 0.184$ corresponds to the luminosity distance of $d_{\rm L} = 884$~Mpc.

\section{Observational data: PDS~456} \label{sec:pds}

\begin{table}
\caption{{\it Suzaku} observations of PDS~456}
\centering
\begin{tabular}{llc}
\hline\hline
Obs ID & Start Date & Net exposure (ks)\\
\hline
701056010 & 2007-02-24 17:58:04 & 190.6\\
705041010 & 2011-03-16 15:00:40 & 125.5\\
707035010 & 2013-02-21 21:22:40 & 182.3\\
707035020 & 2013-03-03 19:43:06 & 164.8\\
707035030 & 2013-03-08 12:00:13 & 108.3\\
\hline
\end{tabular}
\label{suzakuobs}
\end{table}%

PDS~456 has been observed between 2007 and 2013 with {\it Suzaku} \citep{Mitsuda2007}, for a total of five epochs as summarised in Table \ref{suzakuobs}.
Among these observations we choose the 2007 data, as this has strong wind absorption lines from H- and He-like iron. 
It also has a steep spectrum with very little absorption from 
lower ionisation species as required by our code (see \citealt{Reeves2009,Reeves2014}).

We processed and screened XIS data by running {\sc aepipeline} and applied default data screening and cleaning criteria: 
grade 0, 2, 3, 4 and 6 events were used, while hot and flickering pixels were removed,
data were excluded within 436~s of passage through the South Atlantic Anomaly (SAA),
and within an Earth elevation angle (ELV) $<5^\circ$ and Earth day-time elevation angles (DYE\_ELV) $<20^\circ$.
The total net exposure time is 190.6 ks.
Spectra were extracted from circular regions of 2.$\arcmin$9 diameter, while background spectra were extracted from annular region from 7.$\arcmin$0 to 15.$\arcmin$0 diameter.
We generated the corresponding response matrix (RMF) and auxiliary response (ARF) files by utilizing {\sc xisrmfgen} and {\sc xissimarfgen}.
The spectra and response files for the two front-illuminated XIS 0 and XIS 3 chips were combined using the ftool {\sc addascaspec}.
The XIS spectra were subsequently grouped to HWHM XIS resolution of $\sim0.075$~keV at 5.9~keV and $\sim0.020$~keV at 0.65~keV, and then grouped to obtain a minimum 40 counts in each bin.  

Since the main interest of this paper is emission and absorption
feature from the H- and He-like iron, we ignore the spectrum below
2~keV (observed frame) to exclude the soft excess.  We assume that the
2-10~keV continuum can be modelled by a power law over this restricted
energy band, with column density fixed to the Galactic value of
$2\times 10^{21}$~cm$^{-2}$. In the remainder of this section we use
phenomenological models for the absorption and emission at iron to
connect to previous studies. We then use these to estimate the input
parameters for Monte-Carlo simulations of the wind (Section \ref{sec:param}).  We
show all spectra in the rest frame of PDS~456.

\subsection{Gaussian absorption and emission}\label{sec:2007gaus}

We fit two negative Gaussian lines to characterize the absorption, plus a
single positive Gaussian line to characterise the emission atop a power law continuum. The equivalent
width of the absorption lines is $0.110^{+0.035}_{-0.028}$ (He-like) and $0.094_{-0.035}^{+0.025}$~keV
(H-like).  We confirm the results of \cite{Reeves2009} that the
He-like and H-like absorption features have slightly but significantly
different blueshift, at $v_{out}=0.295\pm 0.005c$ (He-like) and
$0.310\pm 0.007c$ (H-like). The two absorption lines are constrained
to have the same intrinsic width, which is marginally resolved
($\sigma=0.048(<0.096)$~keV). By constrast, the emission line is
extremely broad, with $\sigma=1.3_{-0.6}^{+1.5}$~keV and equivalent width
$0.35_{-0.28}^{+0.28}$~keV.  The power law continuum is quite steep at
$\Gamma=2.34_{-0.05}^{+0.10}$, and this is a good fit overall, with
$\chi^2=99.33/98$. All parameters are listed in Table \ref{param}.

\begin{figure}
\centering
\includegraphics[width=8cm]{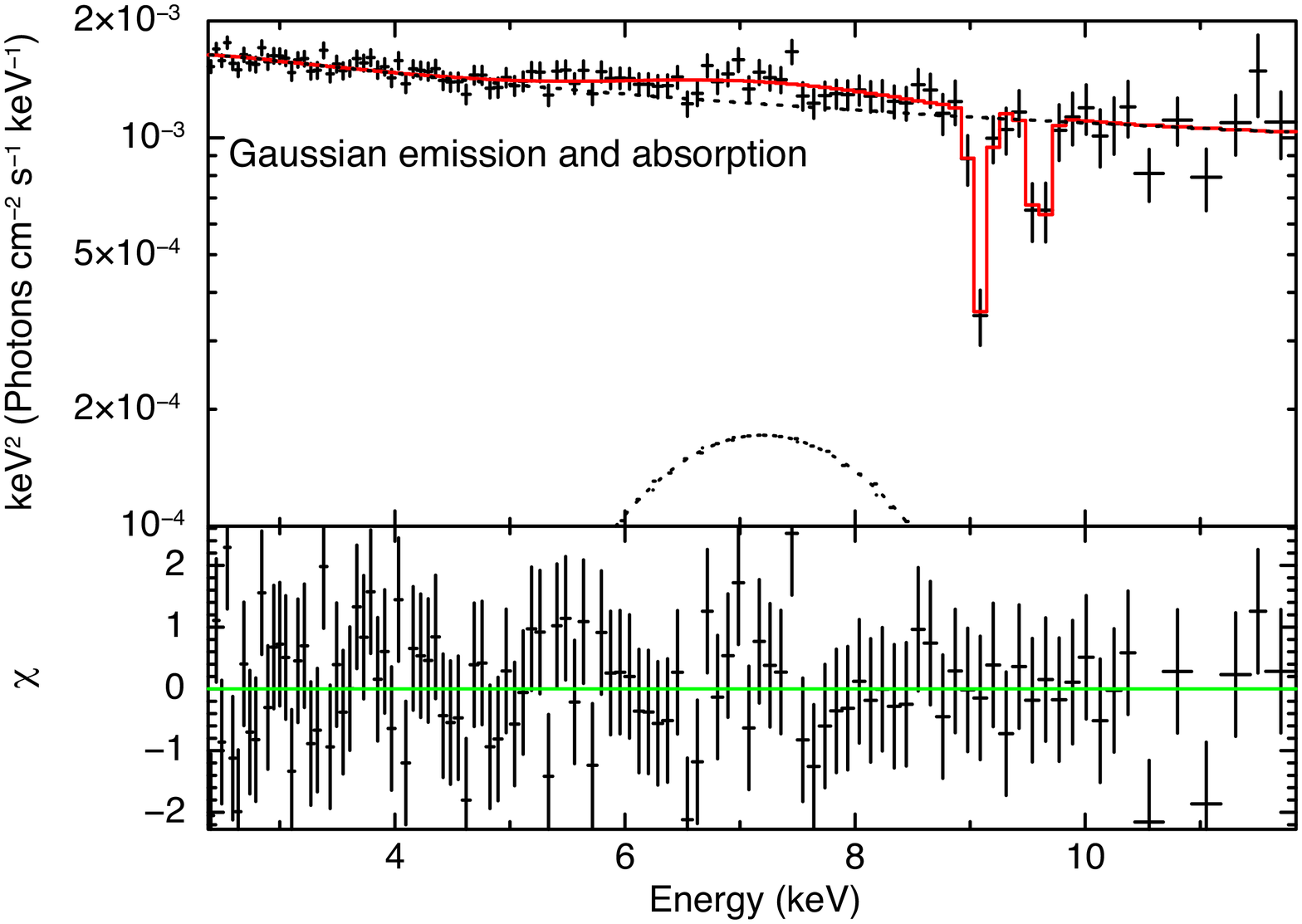}
\includegraphics[width=8cm]{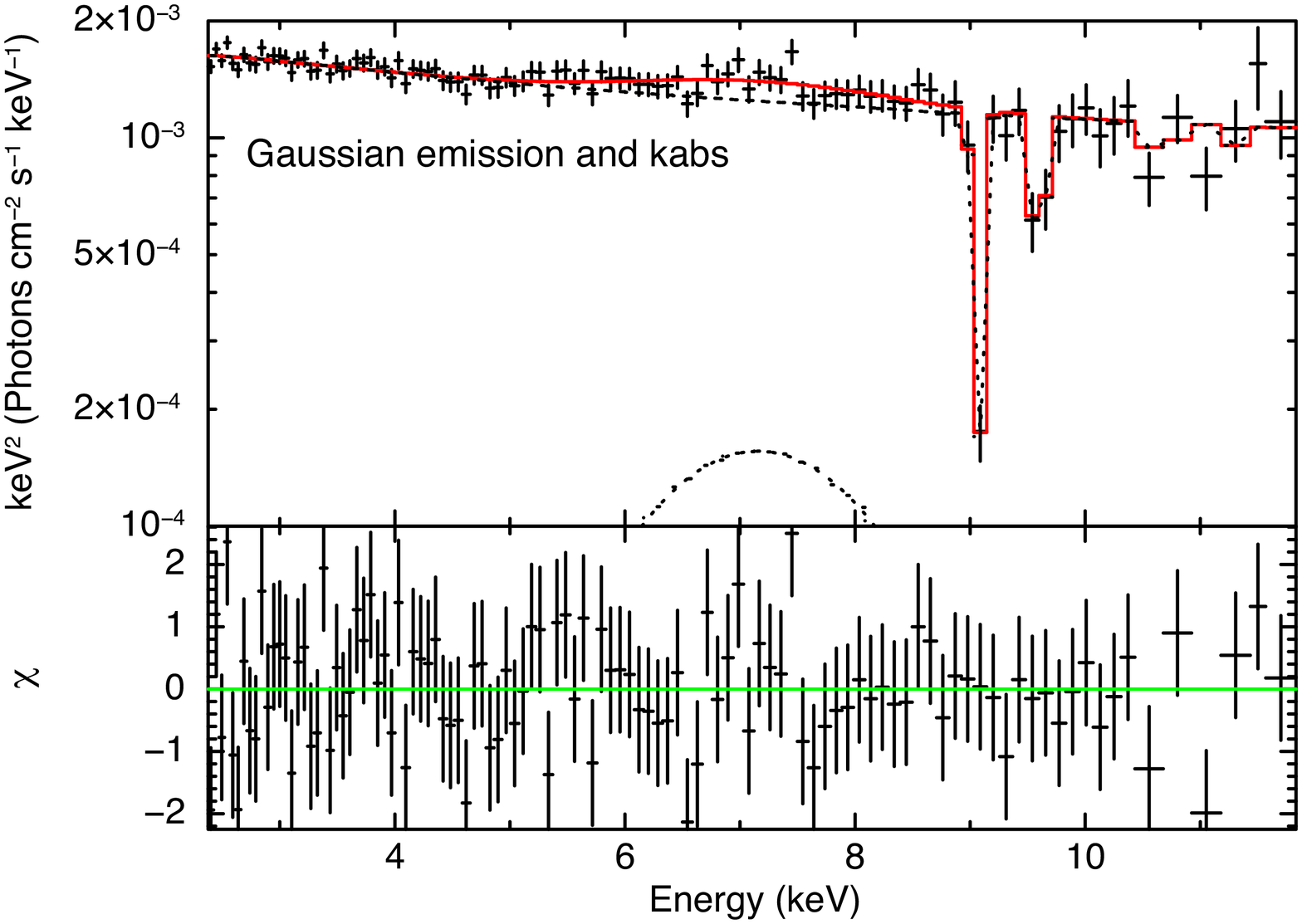}
\includegraphics[width=8cm]{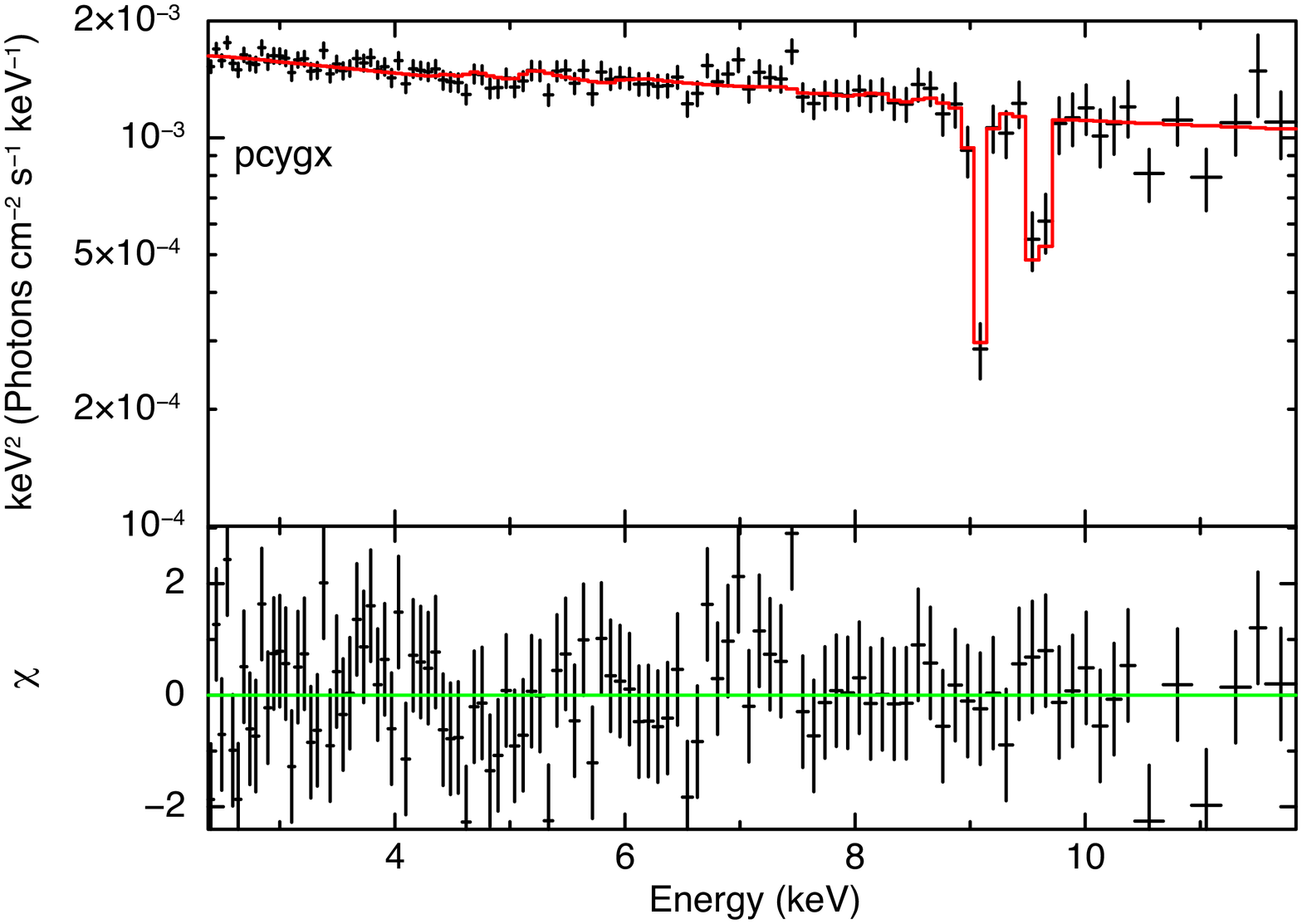}
\caption{{\it Suzaku} spectra fitted with different parameters.
Top : Gaussian emission and absorption, Middle : {\sc kabs} and Gaussian emission, Bottom : {\sc pcygx}.
The spectra are shown in the rest frame of PDS~456.
}
\label{default}
\end{figure}

\begin{table}
\caption{Spectral parameters for the 2007 spectrum}
\begin{center}
\begin{tabular}{llc}
\hline\hline
Model Component & Fit Parameter & Value (90\% error)\\
\hline
\multicolumn{3}{c}{Gaussian absorption and emission}\\
\hline
Powerlaw	& $\Gamma$ & $2.34^{+0.10}_{-0.05}$\\
		& $F_\mathrm{2-10~keV}$ ($10^{-12}$~erg~s$^{-1}$~cm$^{-2}$) & $3.77^{+0.07}_{-0.22}$\\
		& $L_\mathrm{2-10~keV}$ ($10^{44}$erg~s$^{-1}$) & $3.52^{+0.07}_{-0.21}$\\
FeXXV	&$v_{out}$& $0.295^{+0.005}_{+0.005}c$\\
(6.6975 keV)& $\sigma$ (keV) & $0.048 (<0.096)$\\
		& EW (keV) & $0.110^{+0.035}_{-0.028}$\\
FeXXVI	&$v_{out}$& $0.310^{+0.007}_{-0.007}c$\\
(6.9661 keV)& $\sigma$ & tied to FeXXV\\
		& EW (keV) & $0.094^{+0.025}_{-0.035}$\\
Emission	& LineE (keV) & $6.74^{+0.47}_{-1.32}$\\
		& $\sigma$ (keV) & $1.27^{+1.48}_{-0.59}$\\
		& EW (keV) & $0.35_{-0.28}^{+0.28}$\\
Fit statistics	     & $\chi^2$/dof & 93.33/98\\
			     & Null probability & 0.61\\
			     & $\chi^2$/dof for 6.5--10.0~keV & 13.90/20\\
\hline
\multicolumn{3}{c}{{\sc kabs} + Gaussian emission}\\
\hline
Powerlaw	& $\Gamma$ & $2.32_{-0.05}^{+0.06}$\\
		& $F_\mathrm{2-10~keV}$ ($10^{-12}$~erg~s$^{-1}$~cm$^{-2}$) & $3.78_{-0.12}^{+0.11}$ \\
		& $L_\mathrm{2-10~keV}$ ($10^{44}$~erg~s$^{-1}$) & $3.53_{-0.11}^{+0.10}$ \\
FeXXV 	&$v_{out}$& $0.294_{-0.004}^{+0.004}c$\\
	    	& kT (keV) & $474(<10484)$\\
		& Natom ($10^{18}$) & $3.7_{-2.1}^{+18.9}$\\
		& EW (keV) & $0.122$\\
FeXXVI 	&$v_{out}$& $0.310_{-0.006}^{+0.007}c$\\
	    	& kT (keV) & tied to FeXXV\\
		& Natom ($10^{18}$) & $3.0_{-1.7}^{+8.1}$\\
		& EW (keV) & $0.097$\\
Emission	& LineE (keV) & $6.8_{-0.6}^{+0.4}$ \\
		& $\sigma$ (keV) & $1.1_{-0.9}^{+0.9}$ \\
		& EW (keV) & $0.271_{-0.182}^{+0.199}$\\
Fit statistics	     & $\chi^2$/dof & 91.74/98\\
			     & Null probability & 0.66\\
			     & $\chi^2$/dof for 6.5--10.0~keV & 13.09/20\\
\hline
\multicolumn{3}{c}{{\sc pcygx}}\\
\hline
Powerlaw	& $\Gamma$ & $2.37_{-0.03}^{+0.04}$\\
		& $F_\mathrm{2-10~keV}$ ($10^{-12}$~erg~s$^{-1}$~cm$^{-2}$) & $3.79_{-0.05}^{+0.05}$ \\
		& $L_\mathrm{2-10~keV}$ ($10^{44}$~erg~s$^{-1}$) & $3.54_{-0.05}^{+0.05}$ \\
FeXXV	&$v_{out}$& $0.356_{-0.006}^{+0.007}c$\\
(6.6975 keV)& $\tau_{tot}$ & $0.018_{-0.017}^{+6.577}$ \\
		& $\alpha$ & $-10.8_{-0.2}^{+1.6}$\\
FeXXVI	&$v_{out}$& $0.378_{-0.009}^{+0.009}c$\\
(6.9661 keV)& $\tau_{tot}$ & $0.010(<2.544)$\\
		& $\alpha$ & tied to FeXXV\\
Fit statistics	     & $\chi^2$/dof & 102.24/101\\
			     & Null probability & 0.45\\
			     & $\chi^2$/dof for 6.5--10.0~keV & 15.36/23\\
\hline
\end{tabular}
\end{center}
\label{param}
\end{table}%


\subsection{Physical absorption lines: {\sc kabs} plus Gaussian emission}
We use a physical absorption line model to estimate physical parameters for the following winds simulations. 
The absorption line profile should be a combination of a Gaussian
core, with Lorentzian wings, with the ratio of these two components
depending on the total optical depth of the line transition. This
profile is incorporated in the {\sc kabs} model (\citealt{Kotani2000}
including Erratum in 2006), with the free parameters being the column
density of the ion, together with the temperature (equivalent to a
turbulent velocity). We include FeXXV (He-like) and FeXXVI (H-like)
K$\alpha$ and $\beta$, so have 4 absorption lines, but we note that
the K$\beta$ lines are determined self consistently from the K$\alpha$
line parameters so the fit has the same number of free parameters as the fit with two lines.

This gives an equivalently good fit, with $\chi^2=91.74/98$. Again the
He-like line velocity is significantly smaller than the H-like, at 
$0.294_{-0.004}^{+0.004}c$ compared to $0.310_{-0.006}^{+0.007}c$. 
The derived line broadening temperature of $\sim 474$~keV 
corresponds to a velocity width 
$\sigma = E_0 (2 kT/[(m_e c^2) (Am_p/m_e)])^{1/2}= 0.029$~keV
i.e. a turbulent velocity of 1300~km~s$^{-1}$, where $A$ is an atomic mass. 

Fixing both ions to this mean turbulence gives a column of FeXXV of
$3.7_{-2.1}^{+18.9}\times 10^{18}$ and of FeXXVI of $3.0_{-1.7}^{+8.1}\times
10^{18}$~cm$^{-2}$.  The ratio is the
important factor in determining the ionisation state, and this gives
$\mathrm{H/He}\sim 0.8 (<3.0)$. It seems most likely that $\mathrm{H/He}\ge 1$ as
otherwise we would expect significant column in FeXXIV and below,
which would result in significant K$\alpha$ absorption lines at lower energies
which are not observed. Fixing $\mathrm{H/He}=2$ gives
$N_H(\mathrm{He})=2.2_{-1.2}^{+4.9}\times 10^{18}$~cm$^{-2}$ and $N_H(\mathrm{H})=4.3_{-2.3}^{+9.9}\times
10^{18}$~cm$^{-2}$. These two ion states give an equivalent H column
is $N_H=(N_{FeXXV}+N_{FeXXVI})/A_{Fe}=2.2\times 10^{23}$~cm$^{-2}$ assuming
$A_{Fe}=3\times 10^{-5}$. This is a lower limit as there can be 
a substantial fraction of material which is fully ionised (FeXXVII),
which produces no absorption lines. 

The strongest line (He-like K$\alpha$) is just saturated despite this
large column as the line velocity width is large. Hence the required
column does not decrease much with increasing velocity. However, there
is a limit to how high the turbulent velocity can be as velocities
larger than 6000~km/s ($\sigma>0.14$~keV, $kT>10000$~keV) give lines which are broader
than observed. This forms a lower limit to the He-like and H-like
columns of $1.8$ and $2.5\times 10^{18}$~cm$^{-2}$, respectively.
Decreasing the velocity mean both K$\alpha$ lines saturate, so the
column increases strongly.  The lines are marginally resolved in the
data, but the profiles are heavily saturated at very low line widths
so the lines are broad despite the Doppler core being narrow. Thus
there is no formal lower limit to the velocity. However, the gas is
highly ionised so is also heated to the local Compton temperature
which must be of order $10^6$~K ($kT\sim0.1$~keV). This fixes the upper limit to the
column in He and H-like ions of $220$ and $270\times
10^{18}$~cm$^{-2}$. This would be Compton thick, with $N_H>1.6\times
10^{25}$~cm$^{-2}$.


\subsection{Absorption plus emission: {\sc pcygx}}

The very broad emission line obtained by the above analysis could be
produced by reflection from the disc, but some part of it should also
be produced by the same wind structure that produces the absorption
lines. We can estimate the maximum emission that could be produced by
the wind by using the P Cygni
profile code from \cite{Lamers1987}, as
incorporated into {\sc xspec} by \cite{Done2007}.
This code was
designed to model O star winds, i.e. a spherically symetric, radial
outflow. This clearly differs from the discwind geometry envisaged
here, where the wind is not spherical and the velocity structure includes
rotation as well as radial outflow. However, it gives a zeroth order 
estimate of the strength of emission which might be produced. 

The optical depth in each transition is parameterised as
$\tau(w)=\tau_{tot}(1-w)^{\alpha}$ where
\begin{eqnarray}
w(r)=v(r)/v_{\infty}=w_0+(1-w_0)[1-(r_{launch}/r)]^\beta
\end{eqnarray}
where $w_0$
is the initial velocity of the material, $\beta$ is a parameter
determining the acceleration, which is fixed at 1. 
This equation is an extension of the classical CAK model \citep{Castor1975}, which approximately obey this with $\beta=0.5$.
This gives $v_\infty=1.05(1.09)\times
10^5$~km~s$^{-1}$ for He(H)-like, with a very steep $\alpha\sim -10$
(tied between both ions) and so that most of the column is at $v\sim
v_\infty$.  Hence most of the emission is also concentrated at
velocity $v\sim v_{\infty}$ but is produced at all azimuths, so the
projected velocity (which sets the red and blueshifts) ranges from
$-v_\infty \to v_\infty$ i.e. from 4.9-9.1~keV (He-like) and 
5.1-9.5~keV (H-like).

The increase in $\chi^2$ from 91.7/98 in {\sc kabs} to 102.2/101 in
{\sc pcygx} is significant at less than 99\% confidence as there are 3
fewer degree of freedom (the emission line energy, width and
intensity), so $F=\Delta \chi^2/\Delta \mathrm{dof} = 10.5/3 = 3.5$.
This shows that the observed broad emission is consistent with arising
from the wind rather than requiring a substantial contribution from
reflection from the disc.

\section{Monte-Carlo simulations of the wind} \label{sec:code}

\subsection{Model setup}

In order to synthesize the spectrum from the ionised wind efficiently, we separately perform the calculation of the ionisation structure and the radiative transfer simulation. In the first step, we determine the ionisation structure, i.e. spatial distribution of the ion fractions and the electron temperature, by considering ionisation and thermal balances when one-dimensional radiative transfer from the central source is assumed for simplicity. For this calculation we use XSTAR \citep{Kallman2004}. Once the ionisation structure is obtained, we then perform detailed three-dimensional radiative transfer simulation which treats the Doppler effect due to gas motion and photon transport in a complicated geometry. This calculation procedure was established in the context of X-ray spectral modeling of a photoionised stellar wind in a high-mass X-ray binary \citep{Watanabe2006}.

\subsubsection{Geometry} \label{sec:geometry}

\begin{figure}
\begin{center}
\includegraphics[width=8cm]{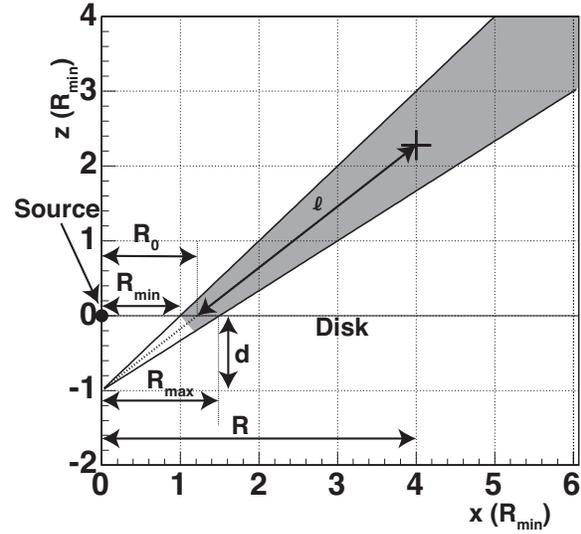}
\caption{Adopted geometry for our wind model and geometric variables. The shaded region is filled with outflowing materials. The wind is rotational symmetrical about z-axis.}
\label{geometry}
\end{center}
\end{figure}

We follow the geometry of \cite{Sim2008,Sim2010a}, where a
biconical configuration as shown in Fig.\ref{geometry} is adopted.  This
geometry was developed for studying radiative transfer in the wind of
cataclysmic variables \citep{Shlosman1993} and widely used for
accretion disc winds \citep{Knigge1995,Sim2008,Sim2010a}.

This geometry is defined by 3 parameters. All stream lines in the wind
converge at a focal point, which is at a distance $d$ below the
source. The wind is lauched from $R_{min}$ to $R_{max}$ on the disc,
We first assume that $d=R_{min}$ and $R_{max}=1.5R_{min}$. This
means that the wind fills a bicone
between $\theta_{min}=45^\circ$ and
$\theta_{max}=56.3^\circ$ i.e. has solid angle $\Omega/4\pi=0.15$.  

We define a mean launch radius $R_0$ from the mean streamline
i.e. it makes an angle of
$\theta_0\equiv(\theta_{min}+\theta_{max})/2$. Thus,
$R_0=d\tan\theta_0$.
The outer boundary of the wind geometry is assumed to be $5\times10^{18}$~cm. As
described in following sections, we chose
$R_{min}=20R_g\simeq 5.9\times10^{15}$~cm for
$M_{BH}=2\times10^9$~$\mathrm{M_\odot}$\citep{Reeves2009}.
Therefore, the outer boundary is three order of magnitude larger
than an inner radius $R_{min}$. Thus, the density at the outer boundary is negligible compared with that at $R_{min}$.
The geometry is divided into 100
shells. Each shell has an equal width on a logarithmic
scale. The radial and azimuthal velocity is assigned at a center on a
logarithmic scale for each shell.

\subsubsection{Velocity and mass}

Radial velocity is defined as a function of length along the streamline $l$
\begin{eqnarray}
v_r(l) = v_0 + (v_{\infty} - v_0) \left( 1 - \frac{R_{min}}{R_{min} + l} \right) ^{\beta}.
\end{eqnarray}
$\beta$ determines the wind acceleration law, similarly to the P Cygni
wind profile in section 2.3, while $v_0$ and $v_\infty$ are an initial
radial velocity at $l=0$ and radial velocity at $l=\infty$.  The azimuthal
velocity at the launching point $R_{0}$ is assumed to be the Keplerian
velocity $v_{\phi_0} = \sqrt{GM/R_{0}}$. 

According to angular
momentum conservation, $v_\phi$ is written as a function of $R$
\begin{eqnarray}
v_\phi(R) = v_{\phi_0} \frac{R_{0}}{R}.
\end{eqnarray}
The turbulent velocity $v_{turb}$ is composed of intrinsic turbulent velocity $v_{t}$ and velocity shear (Appendix A4 of \citealt{Schurch2007}).
\begin{eqnarray}
v_{turb}(i) = v_{t} + \frac{v_r(i)-v_r(i-1)}{\sqrt{12}},\label{vturb} 
\end{eqnarray}
where index $i$ refers to the shell number and $v_r$ is a radial velocity.

According to mass conservation, total mass outflow rate $\dot{M}_{wind}$ is constant.
Therefore, density $n$ is written as 
\begin{eqnarray}
\dot{M}_{wind} &=& 1.23m_p n v_r 4\pi D^2 \frac{\Omega}{4\pi}\label{mass0}\\
		     &=& 1.23m_p n v_r 4\pi D^2 (\cos\theta_{min}-\cos\theta_{max}).\label{mass}
\end{eqnarray}
Here, $D=R/\sin\theta_0$ is the distance from the
focal point, $1.23m_p$ is an ion mass and $\Omega$ is the solid angle of the wind including both sides of the disc.

\subsubsection{Ionisation calculation}

\begin{figure}
\begin{center}
\includegraphics[width=8cm]{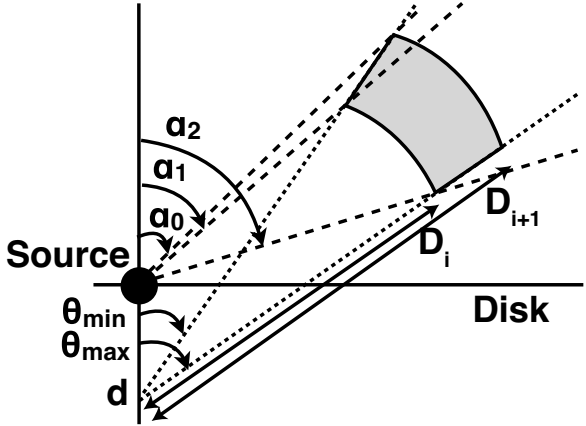}
\caption{A geometry used for ionisation calculation.}
\label{XSTARcalc}
\end{center}
\end{figure}

We run XSTAR version 2.2.1bn16 to calculate the ionisation structure sequentially from the inner shell to the outer shell.
From the output spectrum of each XSTAR run we calculate an input spectrum for the next shell.
For each shell, there are two kinds of input photons. One is the photons directly come from the source, the other is those transmitted and emitted outward.
The fraction of the directly incident component can be calculated geometrically (Fig. \ref{XSTARcalc}).

\begin{eqnarray}
\alpha_0(i) &=& \arctan\left( \frac{D_{i+1}\sin\theta_{min}}{D_{i+1}\cos\theta_{min}-d} \right)\\
\alpha_1(i) &=& \arctan\left( \frac{D_i\sin\theta_{min}}{D_i\cos\theta_{min}-d} \right)\\
\alpha_2(i) &=& \arctan\left( \frac{D_i\sin\theta_{max}}{D_i\cos\theta_{max}-d} \right)\\
f_{direct} &=& (\alpha_1 - \alpha_0)/(\alpha_2 - \alpha_1)\label{f_direct}
\end{eqnarray}
$f_{direct}$ of source photon incident directly, and $1-f_{direct}$ of transmitted and outward-emitted photons become an input spectrum of the next shell.

An input spectrum for XSTAR should be defined in 1--1000~Ry (0.0136--13.6~keV) energy band. Although we don't know PDS~456 spectrum in the UV band, we extrapolate a simple powerlaw with photon index $\Gamma=2.5$. The ionisation luminosity in this energy range is calculated from 2--10 keV X-ray luminosity. If we use $\Gamma=2.2$, the ionisation parameter $\log\xi$ decreases by 10--20\%.

XSTAR requires density $n$, luminosity $L$ and ionisation parameter $\log\xi=\log(L/(nR^2))$ for input parameters.
The density and luminosity are calculated by Eq.\ref{mass} and Eq.\ref{f_direct}.
To get the ionisation parameter, the distance $R$ is needed. Here, the distance $R$ is defined to be a distance between the source and the inner edge of each shell.
Additionally, we inputted a turbulent velocity calculated by Eq. \ref{vturb}.
Atomic abundances are assumed to be equal to the solar abundances for all elements.

\subsubsection{3-dimensional radiative transfer}
We use our Monte-Carlo simulation code called MONACO \citep{Odaka2011} for the detailed radiative transfer.
MONACO is a general-purpose framework for synthesising X-ray radiation from astrophysical objects
by calculating radiative transfer based on the Monte-Carlo approach.
This framework utilises the Geant4 toolkit library \citep{Agostinelli2003, Allison2006} in order to calculate particle trajectories and physical interactions of the particles with matter in a complicated geometry.
MONACO is designed to treat astrophysical applications in which matter can form into an ionised plasma and can have motion that results in the Doppler shifts and broadenings.
A variety of geometries and physical processes of photons are equipped and selectable for different astrophysical applications.

We have already included full treatment of photon processes related to an X-ray photoionised plasma.
Detailed implementation of the physical processes is described in \cite{Watanabe2006}.
The simulation tracks photon interactions with ions, namely photoionisation and photoexcitation; after these interactions reprocessed photons generated via recombination and atomic deexcitation are continue to be tracked.
Compton scattering by free electrons is also taken into account.
In this work, we consider only H- and He-like ions of Fe and Ni, and we ignore other ions.
This assumption is justifiable by the fact that in the region of interest lighter elements are fully stripped, and L-shell ions of Fe and Ni with a few electrons have a small impact on the absorbed spectrum even if they exist.

We divide into 64 parts in azimuthal angle and 2 parts in polar angle
since each cell can have only one velocity vector in this simulation
code. Therefore, $100\mathrm{(radial)}\times
64\mathrm{(azimuthal)}\times 2\mathrm{(polar)}$ cells are constructed
in this Monte-Carlo simulation.
We populate this using a powerlaw spectrum with photon index
$\Gamma=2.5$ in the 5--200~keV energy range.  Initial directions of the seed
photons are limited to the upper half of the disc because photons below
the disc usually cannot penetrate the disc.

\subsection{Parameter choice}\label{sec:param}

We translate the observational data above into appropriate simulation
parameters. Firstly, we assume a minimum turbulent velocity
$v_{turb,0}=10^3$~km~s$^{-1}$ \citep{Reeves2009, Reeves2014}, and set
$v_{\infty}=0.3c$ (maximum velocity of H-like iron)
This implies a launch radius of $R_{min}=20 R_g$ for
$v_{\infty}=v_{esc}=c\sqrt{2R_g/R}$.  We assume that this extends to
$R_{max}=1.5R_{min}=30R_g$. 
We need the wind to be quite likely to intercept our line of sight in order to see absorption, so 
we assumed $\Omega/4\pi=0.15$ \citep{Tombesi2013}.

We assume that the wind is radiation driven, so we
can get some idea of its polar angle from the ratio of luminosity from
$20-30R_g$, which will vertically accelerate the wind, to the
luminosity from $6-20R_g$ which pushes the wind sideways (see
e.g. \citealt{Risaliti2010}; \citealt{Nomura2013}). For a spin zero black hole accreting
at $L=L_{Edd}$ we find $L(20-30R_g)=0.64 L(6-20R_g)$, giving a polar
angle of $\sim 57^\circ$. Hence we choose to fill the solid angle in a
bicone from $45-56.5^\circ$ \citep{Sim2010a, Sim2010b}. 

Conservation of mass (Equation \ref{mass0}) means $n(R)\propto 1/(v_r(R) R^2)$.
The total column density along the wind is $\int_{R_0}^{\infty} n(R)
dR$, so for fast acceleration, where $v(R)\sim v_\infty$ for all $R$
then $\dot{M}_{wind}=4\pi v_{\infty} m_p (\Omega/4\pi) 1.23 N_H
R_{0}$. The lower limit to the total hydrogen column (from the upper
limit to the turbulent velocity) implies
$N_H=(N_{FeXXV}+N_{FeXXVI}+N_{FeXXVII})/A_{Fe}>1.5\times
10^{23}$~cm$^{-2}$ so the absolute minimum mass loss rate is
$\dot{M}_{wind}\sim 0.5$~$\mathrm{M_\odot}$~yr$^{-1}$ for $\Omega/4\pi=0.15$.
Conversely, the upper limit to the column from the lowest velocity
limit implies an upper limit to the mass loss rate of $\sim
50$~$\mathrm{M_\odot}$~yr$^{-1}$, though it could be higher still if there is
substantial material which is completely ionised and hence
invisible. However, these larger columns have very large optical depth
to electron scattering ($\tau_T=1$ corresponds to $N_H=1.5\times
10^{24}$ which corresponds to $5$~$\mathrm{M_\odot}$~yr$^{-1}$), at which point
the wind becomes self shielding, and radiative transfer within the
wind would lead to low ionisation species which are not seen. 
Increasing the mass loss rate increases the optical depth, to $\tau=10$ for $50$~$\mathrm{M_\odot}$~yr$^{-1}$. This would completely obscure the X-ray source along all directions which intercept the wind.

We can set an upper limit on the wind mass loss by the mass accretion rate. 
We use the accretion disc code {\sc optxagnf} \citep{Done2012} with Galactic reddening of 0.48 \citep{Simpson2005} and simulate an accretion disc spectrum for a black hole of mass $2\times 10^9$~$\mathrm{M_\odot}$~yr$^{-1}$. We match the observed B and V band fluxes \citep{Ojha2009} for $L=0.4L_{Edd}$ for a spin 0 black hole, i.e. a bolometric luminosity of $\sim 10^{47}$~ergs~s$^{-1}$ and mass accretion rate of $\dot{M}=31$~$\mathrm{M_\odot}$~yr$^{-1}$. Alternatively, this gives $L=2L_{Edd}$ for a spin 0.998, corresponding to $L_{bol}\sim 5\times 10^{47}$~ergs~s$^{-1}$ and mass accretion rate of $\dot{M}=27$~$\mathrm{M_\odot}$~yr$^{-1}$. 
The lack of dependence of the derived mass accretion rate on black hole spin is as expected, as spin only affects the disc structure on size scales comparable to the last stable orbit, whereas the optical emission which we use to derive mass accretion rate is produced from further out in the disc. 
Clearly the maximum mass loss rate is then equal to the mass input rate of $30$~$\mathrm{M_\odot}$~yr$^{-1}$, but we
set a conservative limit of $15$~$\mathrm{M_\odot}$~yr$^{-1}$, where we can lose up to half of the input mass accretion rate. 

The density of the material is also determined by the opening angle of
the wind with $n(R) \propto \dot{M}_{wind}/[R^2 v(R) (\Omega/4\pi)]$
(Equation \ref{mass0}). A wider opening angle means that the wind is more likely
to intercept the line of sight, but also means that the same mass loss
rate is spread into a larger volume, so this has lower density. This
determines the ionisation parameter $\xi=L/(nR^2) \propto v(R)
(\Omega/4\pi)/\dot{M}_{wind}$, which controls the ratio of H-like to
H-like ion column density. The fact that the data (weakly) require
He-like and H-like to have different velocities implies that the
ionisation is not constant in the wind as might be expected if all the
absorption is produced after the wind has been accelerated to its
terminal velocity (so $v=v_\infty$ and is constant). This shows that
it is feasible to use observational data to constrain the wind
acceleration.

\begin{figure*}
\centering
\includegraphics[width=16cm]{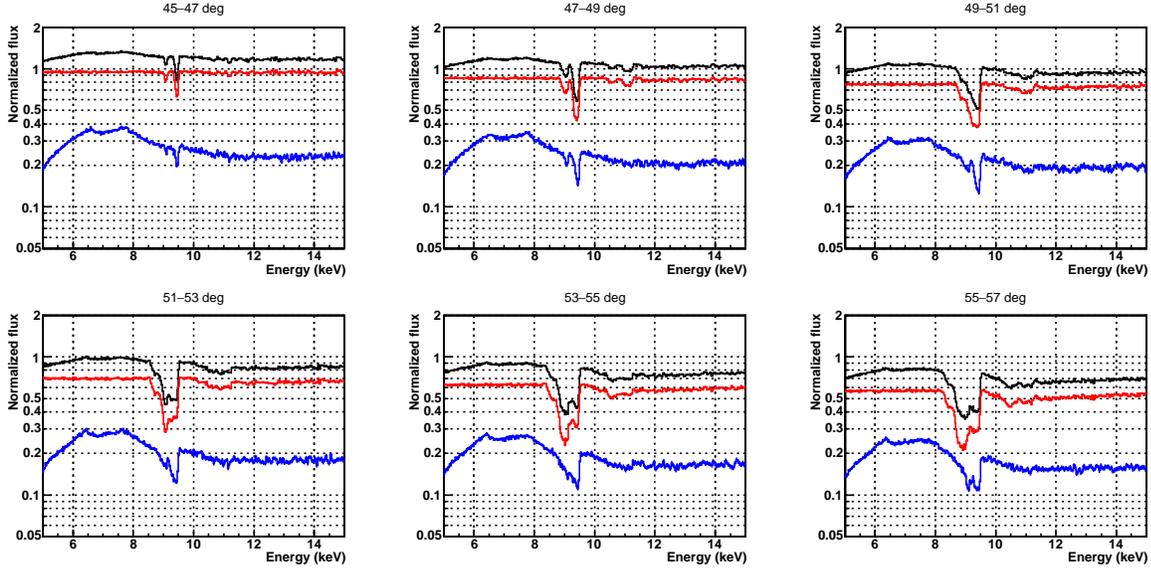}
\caption{MONACO spectra with $L=4\times10^{44}$~erg~s$^{-1}$, $\dot{M}=10$~$\mathrm{M_\odot}$, $v_0=v_{turb}=1000$~km~s$^{-1}$, $\beta=1$ and $R_{min}=20R_g$. The direct component and reprocessed component are plotted in red and blue respectively. The total spectrum is plotted in black. Y-axis is normalised to the input powerlaw spectrum.}
\label{monaco_l4_m10_vt}
\end{figure*}

\begin{figure}
\centering
\includegraphics[width=8cm]{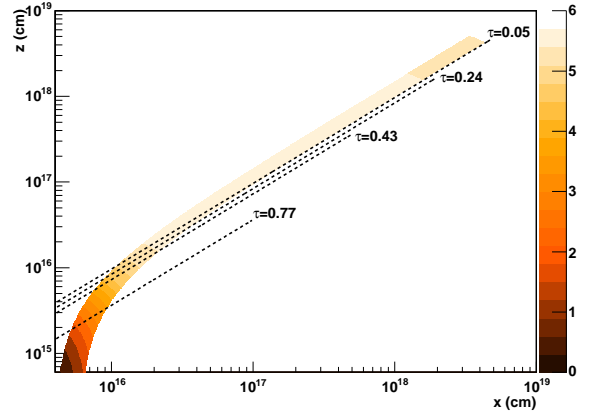}
\caption{Ratio of H-like to He-like iron through the wind, together
with the lines of sight for $\theta_{incl}=46^\circ$, $50^\circ$, $54^\circ$ and $70^\circ$ for the same simulation as in Fig.~\ref{monaco_l4_m10_vt},
labelled with the total column density along that line of sight.
Higher inclination samples material at smaller radii, where it is
still accelerating so the density is higher hence the abundance of
He-like iron is higher.
}
\label{Feratio}
\end{figure}

We show results for $\dot{M}=10$~$\mathrm{M_\odot}$~yr$^{-1}$. We
calculate the ionisation using the measured 2--10~keV X-ray luminosity
of $4\times 10^{44}$~ergs~s$^{-1}$. The results for this for a series
of inclination angles through the wind are shown in Fig.
~\ref{monaco_l4_m10_vt}. The lines clearly increase in both equivalent
width and intrinsic width at higher inclinations, and the ratio of
H-like to He-like iron decreases. 

Fig.~\ref{Feratio} shows the ionisation structure of the wind,
with the lines of sight marked on it.
At larger radii, the product of the density and the radius squared ($nR^2$) is almost constant according to Eq.~\ref{mass0}, due to the saturated velocity. Therefore the H/He ratio shows a slight decrease, which is caused by the decrease of the luminosity due to the wind absorption. On the other hand, since the wind is still accelerating at the smaller radii, there is more He-like than H-like iron (see also \citep{Sim2008}).
As shown in the figure, the high inclination line of sight includes material at 
smaller radii, where the wind is denser and less ionised. This gives the increase in equivalent
width and more He-like than H-like iron.

\section{Comparison of simulations and observations} \label{sec:comp}

\subsection{Absorption lines}\label{sec:comp_abs}
\begin{figure*}
\begin{center}
\includegraphics[width=8cm]{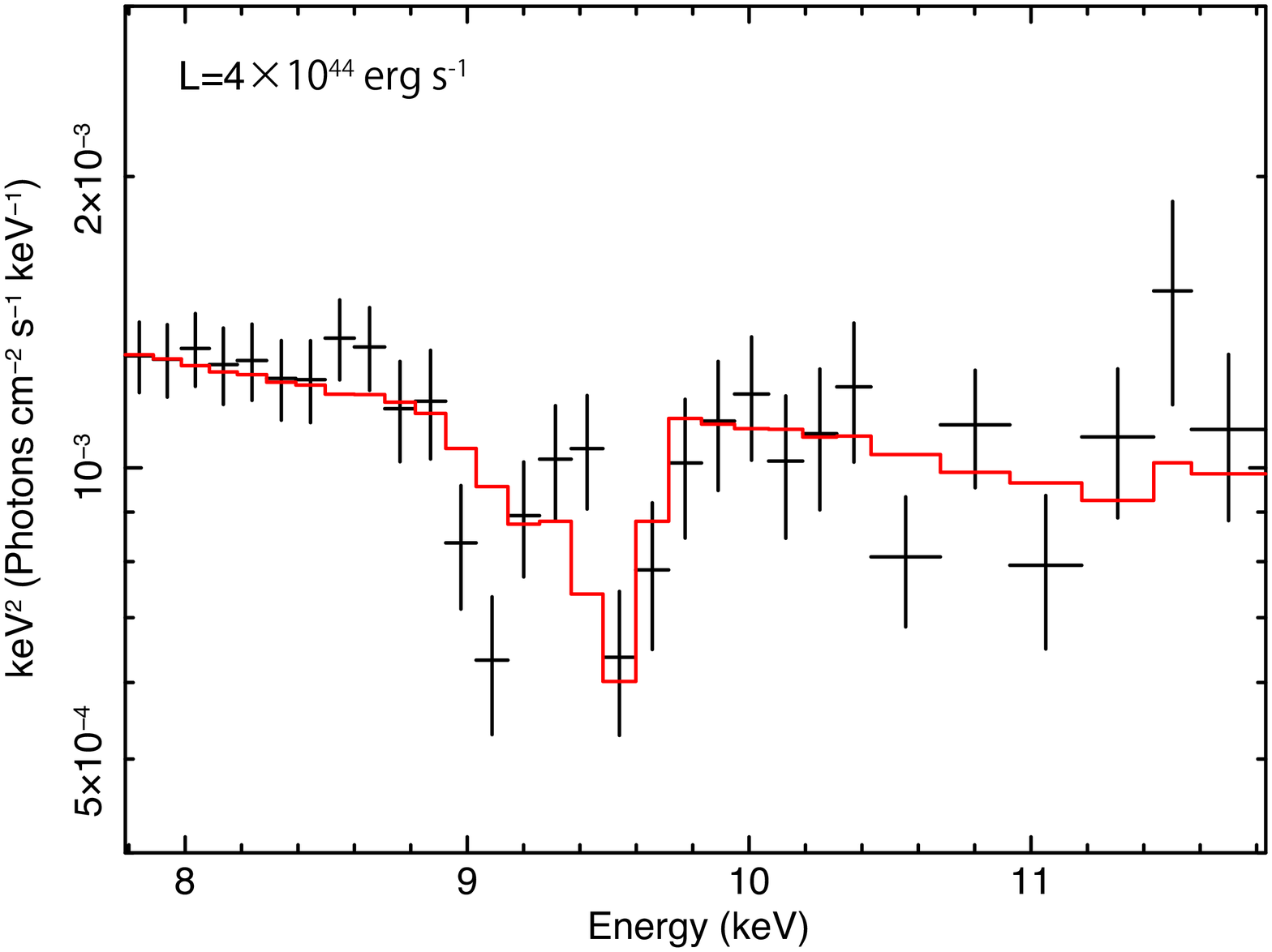}
\includegraphics[width=8cm]{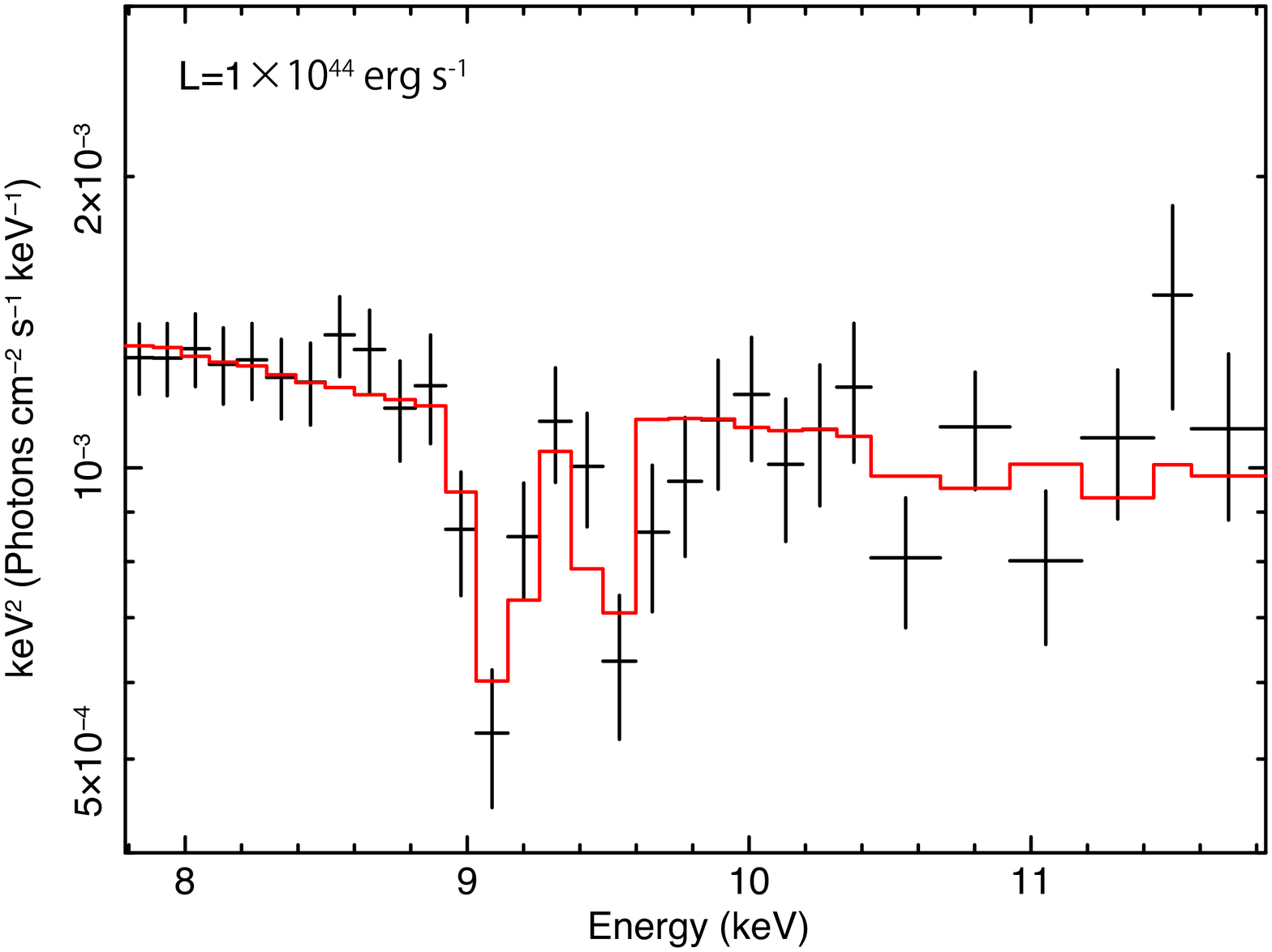}
\caption{{\it Left} : {\it Suzaku} data and MONACO spectrum with $L=4\times10^{44}$~erg~s$^{-1}$, $\dot{M}=10$~$\mathrm{M_\odot}$~yr$^{-1}$, $v_0=v_{turb}=1000$~km~s$^{-1}$, $\beta=1$ and $R_{min}=20R_g$. Best fit parameters are $z=0.165\pm0.007$ ($v\simeq0.315c$), $\theta_{incl}=49.0\pm0.9^\circ$ and $\Gamma=2.35$(fix). Fit statistic is $\chi^2=32.32/27$.
{\it Right} : Same figure as the left panel except with $L=1\times10^{44}$~erg~s$^{-1}$. Best fit parameters are $z=0.174\pm0.005$ ($v\simeq0.308c$), $\theta_{incl}=47.1\pm0.4^\circ$ and $\Gamma=2.35$(fix). Fit statistic is $\chi^2=21.48/27$. All spectra are shown in the rest frame of PDS~456.}
\label{l1_and_l4}
\end{center}
\end{figure*}

The simulation above is close to the largest feasible mass outflow rate, so is close to the
lowest possible ionisation for the observed 2-10~keV X-ray luminosity
of $4\times 10^{44}$~ergs~s$^{-1}$ for the
assumed launch radius of $20-30R_g$
and solid angle $\Omega/4\pi=0.15$. However, it is remarkably difficult
to reproduce the observed absorption line equivalent and intrinsic
widths from this, irrespective of the velocity law chosen, as the
material is very highly ionised (so produces little He-like line)
except at high inclination angles. But at these high inclination
angles, the line of sight intercepts a large range of velocities, so
the lines are broad and blend into each other rather than producing
the two narrow lines seen in the data. Also, material at high
inclination is somewhat shielded from the ionising luminosity by the
rest of the wind. Hence it has lower ionisation state, so at large
inclinations, the He-like ion is produced preferentially at larger
radii than the H-like ion, giving the He-like line a higher outflow
velocity than the H-like, contrary to observations.  Thus both the
narrow line width and the slightly higher velocity in H-like than
He-like imply that the inclination angle through the wind is not too
high, but low inclination angles through the wind are too highly
ionised, producing too small an equivalent width of He-like Fe for
low inclination angles through the wind, and too broad lines
for higher inclination angles. 

We show this by fitting the MONACO model to the 6.5-10 keV data. We tabulate the model as multiplicative factors, and apply these to a power law continuum with Galactic absorption.
The MONACO model has two free parameters of redshift $z$ and inclination angle $\theta_{incl}$. 
We allowed redshift to be free rather than fixing it to the cosmological redshift of $z=0.184$ as this allows us to fit for slightly different wind velocity than is included in the simulation.
The 6.5--10~keV spectrum is used in order to concentrate on the absorption lines. 
The best fit, shown in the left panel of Figure \ref{l1_and_l4}, is not very good, with $\chi^2=32/27$ in the 6.5--10~keV range. This is significantly worse than the phenomenological fits in Table \ref{param}. It is clear from the left panel of Figure \ref{l1_and_l4} that the 
ionisation state of this model is much higher than in the data. 

Changing the velocity law does not substantially change this
conclusion. A much higher initial velocity $v_0=0.15c$ gives a
slightly better fit as this means that the higher inclination lines of
sight through the wind intercept a smaller range of velocity, so the
lines are narrower. Similarly, decreasing $\beta$ also gives a more
homogeneous velocity structure as then most of the acceleration
happens very close to the disc. Full results for these two cases are
shown in the Appendix, but none of these give a particularly good fit
to the data, with $\chi^2>33/27$ for the 6.5-10~keV bandpass.

Increasing the distance at which the wind is launched gives a lower
ionisation parameter. The UV line driven disc wind models of Risaliti
\& Elvis (2010) have $v_\infty /v(R_0)_{esc}\sim $~a few, at which
point the wind could be launched at $R_{min}\sim 50R_g$.  However, the
ratio of accretion power at this point to the total accretion power is
small, so such a wind would be expected to be more equatorial if it is
driven by radiation as the ratio of luminosity under the wind pushing
it up $L(50-75R_g)$ is much smaller compared to $L(6-50R_g)$ which is
the radiation from the inner disc pushing it outwards.

We cannot reduce the ionisation by shielding the gas, as we observe
$L_x=4\times 10^{44}$~ergs~s$^{-1}$ on our line of sight through the
wind, so the wind also should see this luminosity.  However, the
outflow velocity is high enough that the X-ray luminosity as seen in
the rest frame of the wind is substantially reduced by Doppler
de-boosting, so that $L_{obs}=L_x\delta^{3+\alpha}\approx 0.25L_x$
where $\delta=[\gamma(1-\beta\cos\theta)]^{-1}\approx 0.73$ (see
Appendix A3 of Schurch \& Done 2007). Thus the ionising luminosity as
seen by the wind varies from $4-1\times 10^{44}$, depending on the
velocity of the wind. Since the data show that the majority of the
absorption takes place at $v\sim v_\infty$, we use an ionising
luminosity of $10^{44}$~ergs~s$^{-1}$. 

\begin{figure*}
\centering
\includegraphics[width=16cm]{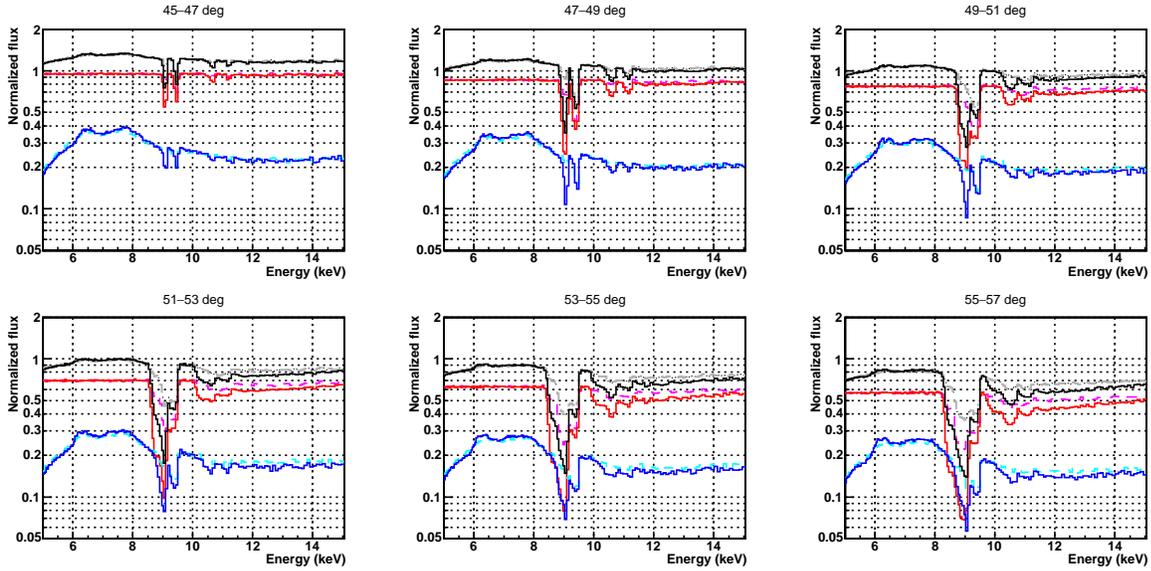}
\caption{Dependence on the ionising luminosity. The grey, magenta and cyan
lines show the fiducial parameter simulation with 
$L=4\times10^{44}$~erg~s$^{-1}$, $\dot{M}=10$~$\mathrm{M_\odot}$~yr$^{-1}$, while the 
black, red and blue dashed curves show the same parameters except with an ionising luminosity $L=1\times10^{44}$~erg~s$^{-1}$.
}
\label{monaco_l1_m10_vt}
\end{figure*}

\begin{figure}
\centering
\includegraphics[width=8cm]{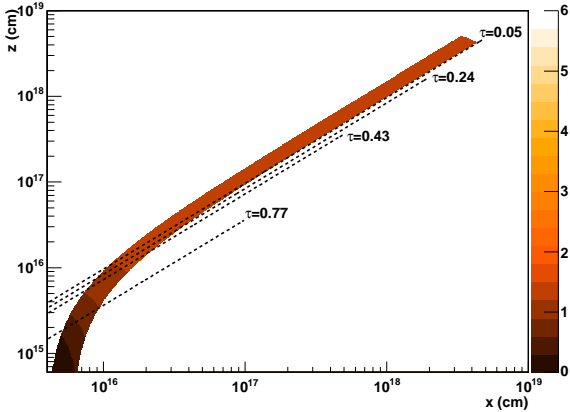}
\caption{Ratio of H-like to He-like iron through the wind, together
with the lines of sight for $\theta_{incl}=46^\circ$, $50^\circ$, $54^\circ$ and $70^\circ$ for the same simulation as in Fig.~\ref{monaco_l1_m10_vt},
labelled with the total column density along that line of sight.
H-like ion is smaller than Fig. \ref{Feratio}.
}
\label{Feratio1e44}
\end{figure}

We re-simulate the wind over a range of parameters with this ionising
luminosity. The full simulation results are shown in
Fig. \ref{monaco_l1_m10_vt} and Fig. \ref{Feratio1e44}, 
showing clearly that the ionisation
state is lower, as expected.

We fit this model to the data, with the best fit 
shown in the right panel of
Fig. \ref{l1_and_l4}. This is a better fit, as expected, 
with fit statistic of 21.5/27, which is not
significantly different to the phenomenological fits in Table
\ref{param}.  We also simulated with
$\dot{M}=15$~$\mathrm{M_\odot}$~yr$^{-1}$,
$8$~$\mathrm{M_\odot}$~yr$^{-1}$, $3$~$\mathrm{M_\odot}$~yr$^{-1}$,
$1$~$\mathrm{M_\odot}$~yr$^{-1}$. Although
$15$~$\mathrm{M_\odot}$~yr$^{-1}$ and $8$~$\mathrm{M_\odot}$~yr$^{-1}$
give comparably good fits, lower wind outflow rates give increasingly
poor fits ($\chi^2=28$ and $51$ respectively) as the absorption lines
become too weak as the material is too highly ionised.

\subsection{Emission lines from the wind}
\begin{figure}
\begin{center}
\includegraphics[angle=-90,width=8cm]{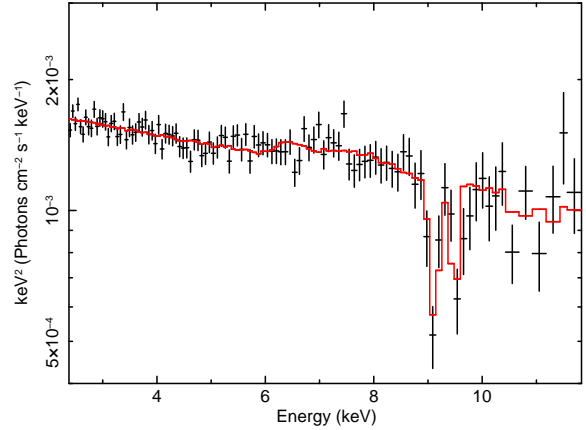}
\caption{{\it Suzaku} data and MONACO spectrum in 2-10~keV band with $L=1\times10^{44}$~erg~s$^{-1}$, $\dot{M}=10$~$\mathrm{M_\odot}$~yr$^{-1}$, $v_0=v_{turb}=1000$~km~s$^{-1}$, $\beta=1$ and $R_{min}=20R_g$. Best fit parameters are $\theta_{incl}=47.3\pm0.4^\circ$ and $\Gamma=2.33\pm0.01$, and the redshift is fixed at $z=0.174$ ($v\simeq0.308c$). Fit statistic is $\chi^2=106.49/105$.}
\label{r20_l1_m10_vt_b1_wide}
\end{center}
\end{figure}

We now re-simulate the best MONACO fit to the absorption lines shown
in the right panel of Fig \ref{l1_and_l4} over an extended energy grid
from 2--200~keV. This enables us to look also at the emission lines
produced by the wind. Fig \ref{r20_l1_m10_vt_b1_wide} shows the best
fit comparison of this simulation with the 2--10~keV Suzaku data,
where the MONACO data are again incorporated as a multiplicative
model. The fit parameters are power law index and normalisation, and
the redshift is fixed at $z=0.174$ ($v_\infty\simeq0.308c$). This
gives $\chi^2=106.5/105$, which is not significantly worse than the
phenomenological fits in Table \ref{param} due to the 
smaller number of free parameters. 
For example, the model using {\sc kabs}
absorption lines with a broad Gaussian emission line has
$\chi^2=91.7/98$, a difference of $\Delta\chi^2=15$ for 7 additional
degrees of freedom. This gives $F=15/7$ which is 2.1, which is only
better at 96\% confidence.

Unlike absorption, the line is emitted from the wind at all azimuths,
and at all radii. Where the wind has already reached its terminal
velocity, it has also expanded enough that its azimuthal velocity is
small compared to its radial outflow velocity. Thus the projected
velocity in our line of sight ranges from $-v_\infty$ ($\theta=0$,
along our line of sight as we look through the wind) to $-v_\infty \cos
(\theta_{incl}+\theta_{max}) \sim -v_\infty \cos2\theta_0$
giving a corresponding line energy of 6.04--9.13~keV for the 6.7~keV
line while the 6.95 keV H-like extends from 6.26-9.47~keV for this simulation. 

Thus the maximum red extension of the emission line can give direct
information on the opening angle of the wind. However, this is
difficult to measure as the line is very broad, and the discussion
above neglects the emission from the wind at small radii where the
initially Keplarian azimuthal velocity is important. 
This line emission from small radii could have a larger projected velocity with $-v_{\phi_0} \cos
(\theta_{incl}+90^\circ)\simeq (v_\infty/\sqrt{2}) \sin(\theta_{incl})$ at maximum, giving a red extension at $\sim5.7$~keV for the He-like line.
Our model shows that the red wing extends down to 6.0--6.3~keV (Fig
\ref{r20_l1_m10_vt_b1_wide}). This would be better matched to the data
if it happened at 6.5--6.7~keV, so we experiment with different
$\theta_{min}$ but keep the same solid angle of the wind. We get
better fit for a wind with $\theta_{min}=35^\circ$ (Fig
\ref{monaco_l1_m10_vt_35deg}) but the decrease in $\chi^2$ is not
significant as these features are all small.

\subsection{Emission lines from the wind and reflection from the disc}

\begin{figure*}
\centering
\includegraphics[angle=-90,width=8cm]{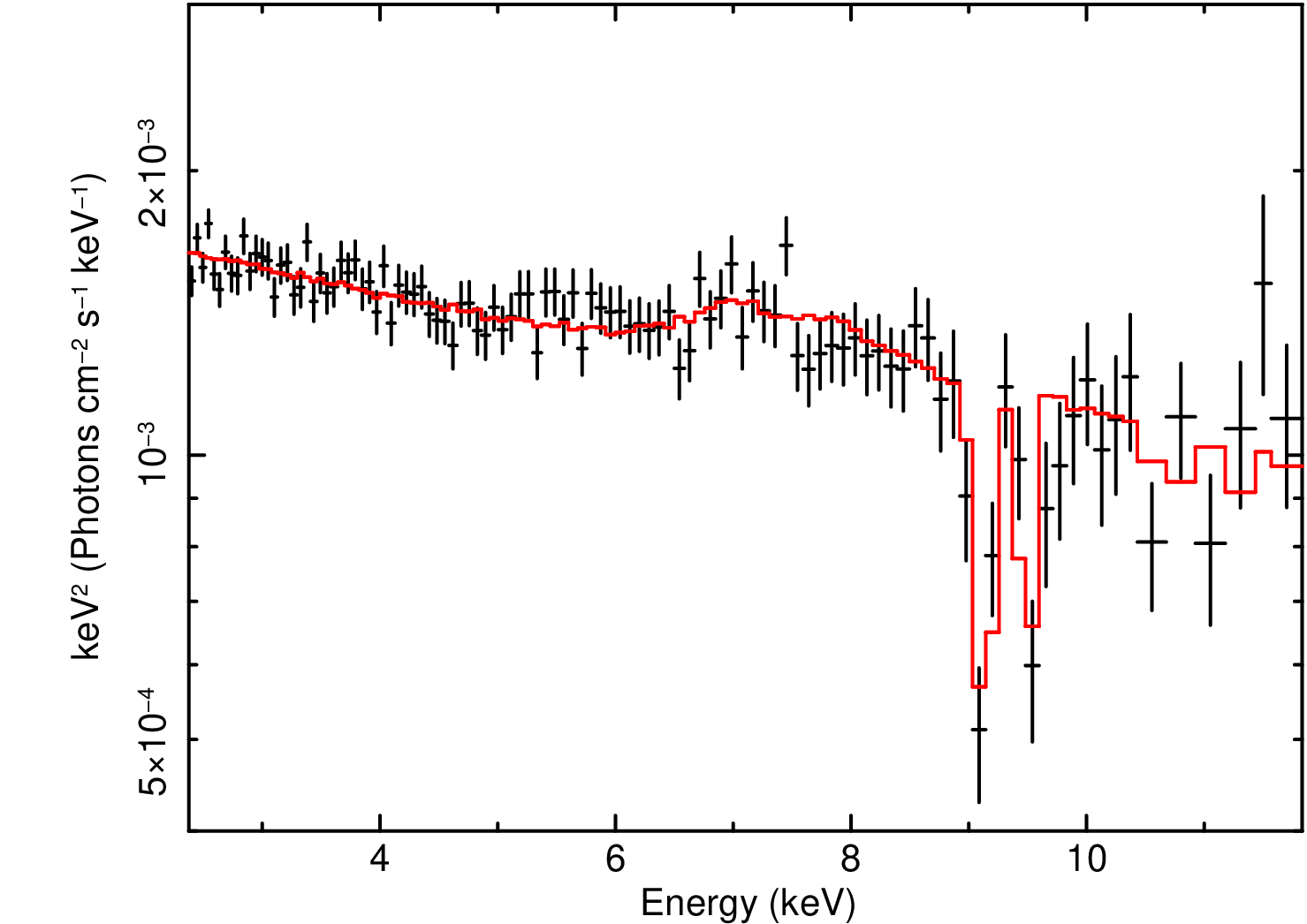}
\includegraphics[angle=-90,width=8cm]{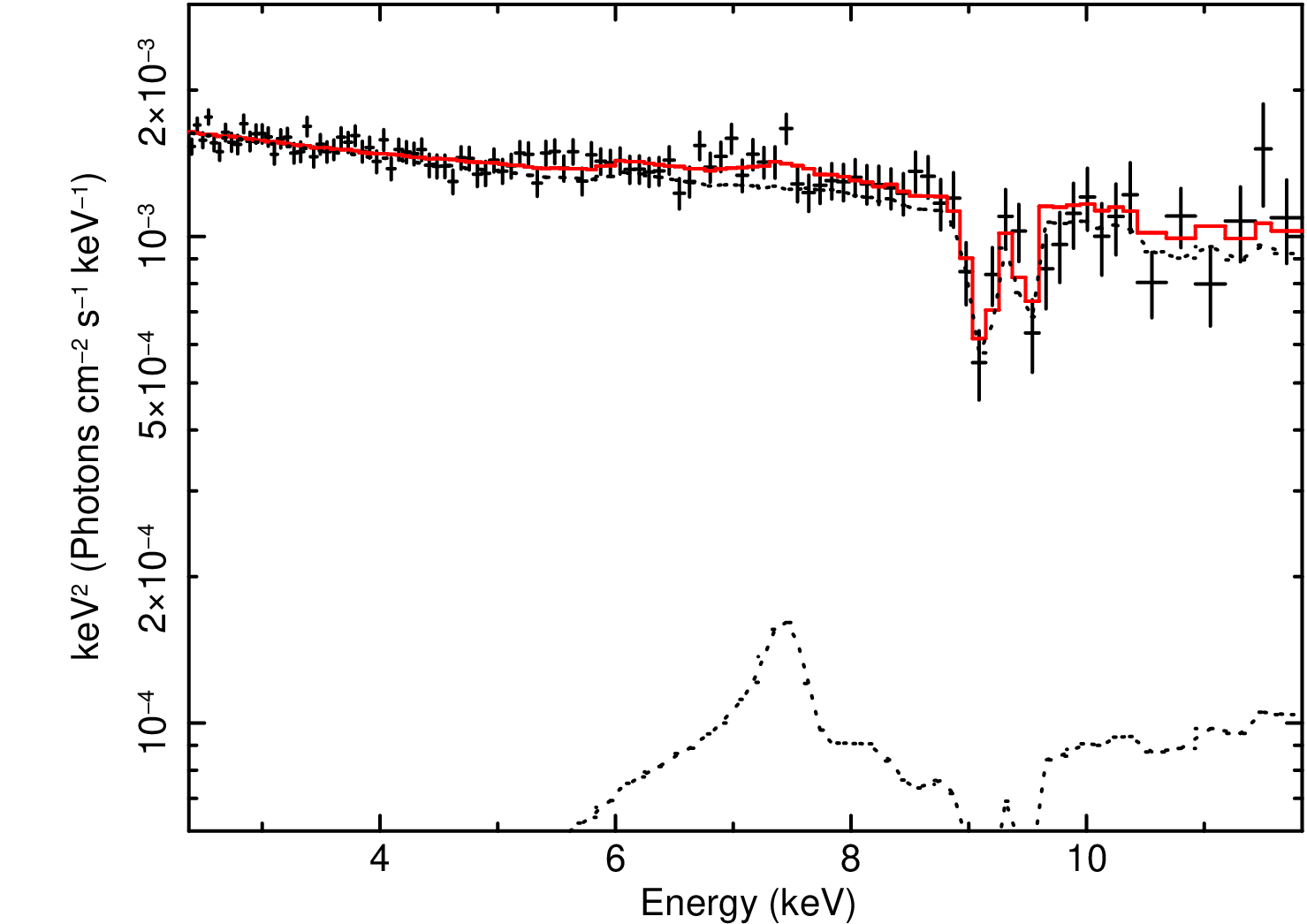}
\caption{{\it Left} : {\it Suzaku} data and MONACO spectrum in 2-10~keV band with $L=1\times10^{44}$~erg~s$^{-1}$, $\dot{M}=10$~$\mathrm{M_\odot}$~yr$^{-1}$, $v_0=v_{turb}=1000$~km~s$^{-1}$, $\beta=1$, $R_{min}=20R_g$ and $\theta_{min}=35^\circ$. Best fit parameters are $\theta_{incl}=37.4\pm0.4^\circ$, $\Gamma=2.30\pm0.01$, and the redshift is fixed at $z=0.174$ ($v\simeq0.308c$). Fit statistic is $\chi^2=103.62/105$. 
{\it Right} : {\it Suzaku} data and MONACO spectrum with blurred disc reflection in 2-10~keV band with $L=1\times10^{44}$~erg~s$^{-1}$, $\dot{M}=10$~$\mathrm{M_\odot}$~yr$^{-1}$, $v_0=v_{turb}=1000$~km~s$^{-1}$, $\beta=1$, $R_{min}=20R_g$ and $\theta_{min}=55^\circ$. Best fit parameters are $\theta_{incl}=56.8\pm0.3^\circ$, $\Gamma=2.40\pm0.04$ and reflection fraction $R\simeq0.27$, and the redshift is fixed at $z=0.174$ ($v\simeq0.308c$). Fit statistic is $\chi^2=103.91/104$. 
We changed the y-axis scale to show the reflected spectrum.
All spectra are shown in the rest frame of PDS~456.
}
\label{monaco_l1_m10_vt_35deg}
\end{figure*}

While the wind produces broadened emission lines from the H- and
He-like material in the wind, the disc should also contribute to the
emission via reflection. In our geometry, the disc still exists from
$20R_g$ down to the innermost stable circular orbit. Hence we include
neutral reflection ({\sc pexmon}) from this inner disc, with
relativistic blurring from {\sc kdblur} with outer radius fixed at
$20R_g$, inner radius fixed at $6R_g$ and emissivity fixed at 3. We
assume that the inclination angle for both {\sc pexmon} and {\sc
kdblur} is tied to inclination angle of the wind model.  We obtained
fit statistics of 103.62, 105.69, 103.91, 107.93 and 111.16, with
reflection fractions of $1\times10^{-3}$, $0.15$, $0.27$, $0.30$ and
$0.35$ for respective values of $\theta_{min}=35$, 45, 55, 65,
75$^\circ$.  We show the fit with $\theta_{min}=55^\circ$ as this
allows a contribution from the inner disc reflection, as expected.
The spectrum is shown in the right panel of
Fig. \ref{monaco_l1_m10_vt_35deg}.

\section{Application to the other observations}\label{sec:otherobs}

We also applied our MONACO models to the {\it Suzaku} data observed on
2011 March 16, 2013 February 21, 2013 March 3 and 2013 March 8 (Table
\ref{suzakuobs}). Hereafter, we refer to the data as 2011, 2013a,
2013b and 2013c respectively.  The data were processed and grouped in
the same way as the 2007 data. The total net exposure times are
125.5~ks, 182.3~ks, 164.8~ks and 108.3~ks respectively.

\begin{figure}
\centering
\includegraphics[angle=-90,width=8cm]{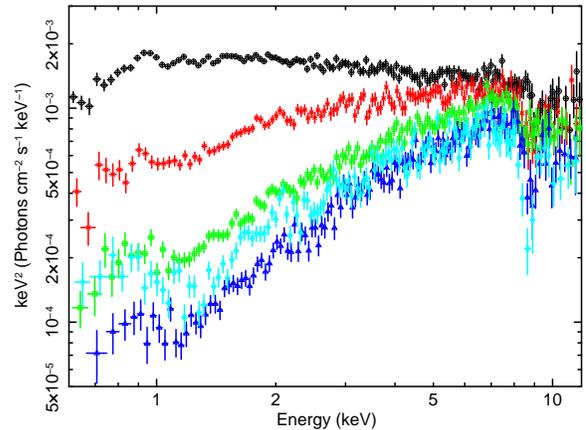}
\caption{{\it Suzaku} XIS FI spectra from 2007 (black circle), 2011 (red cross), 2013a (green square), 2013b (blue diamond) and 2013c (cyan triangle), unfolded against $\Gamma=2$ powerlaw. All spectra are shown in the rest frame of PDS~456.}
\label{allspec}
\end{figure}

Figure \ref{allspec} shows the fluxed spectra of all {\it Suzaku}
observations. The spectra show a large variability in both the
continuum shape and Fe absorption lines. At first sight this
variability appears correlated, with strongest absorption lines in the
hardest spectra. We first assess the extent of the correlation of the
absorption with spectral shape using phenomenological models, and then 
fit using the MONACO spectra. 

\subsection{Spectral fitting with {\sc kabs} model}

\begin{figure*}
\centering
\includegraphics[width=8cm]{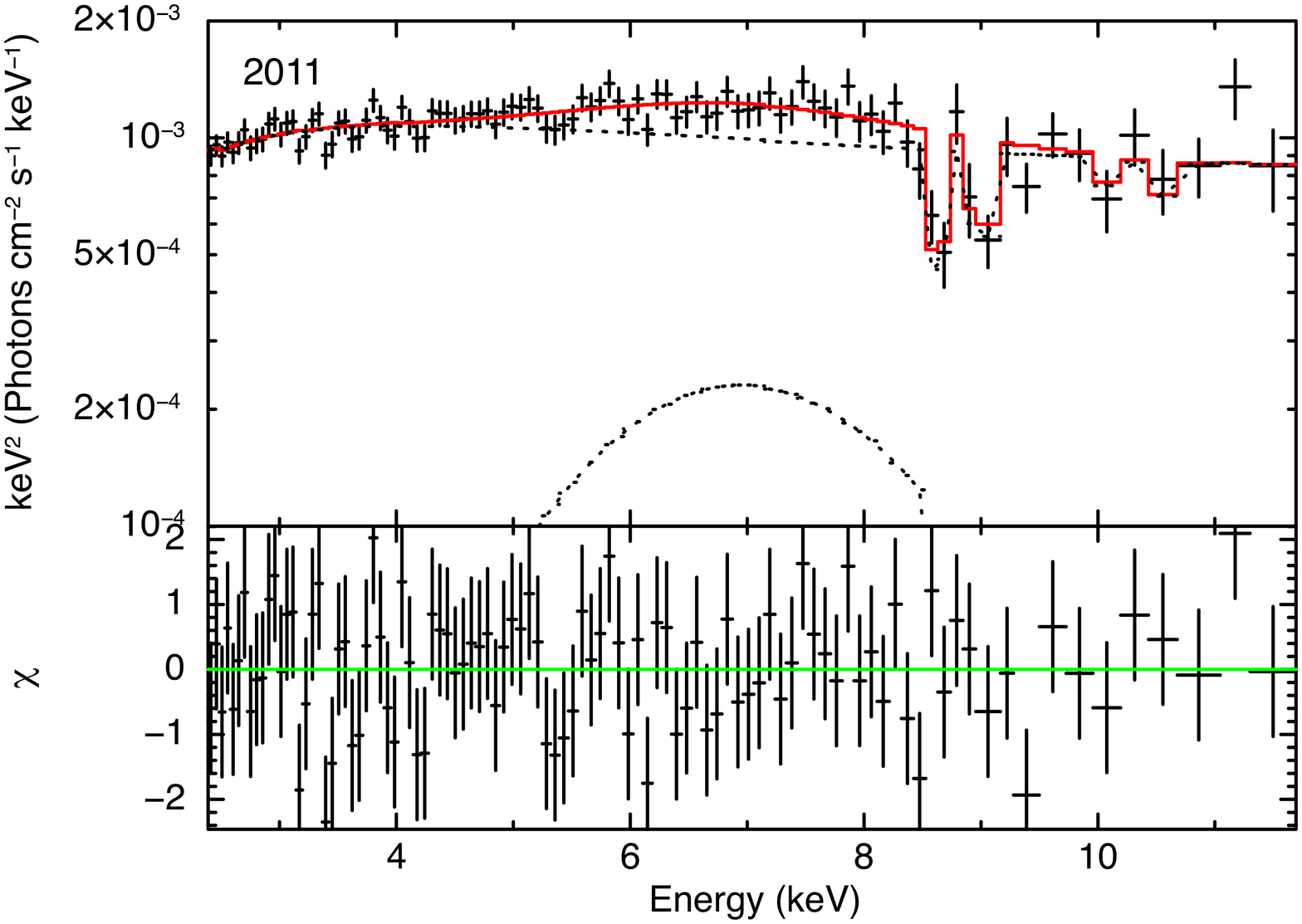}
\includegraphics[width=8cm]{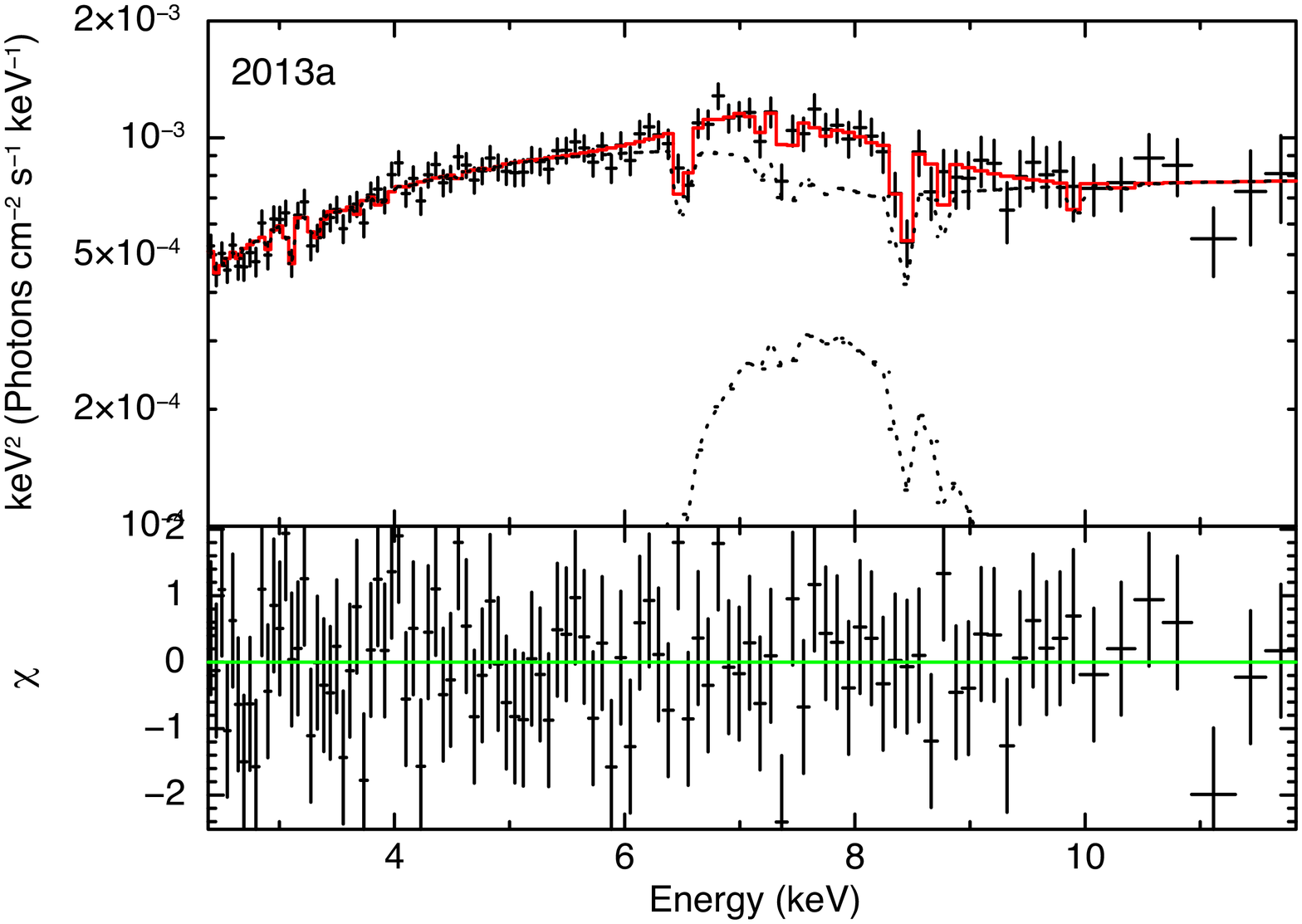}
\includegraphics[width=8cm]{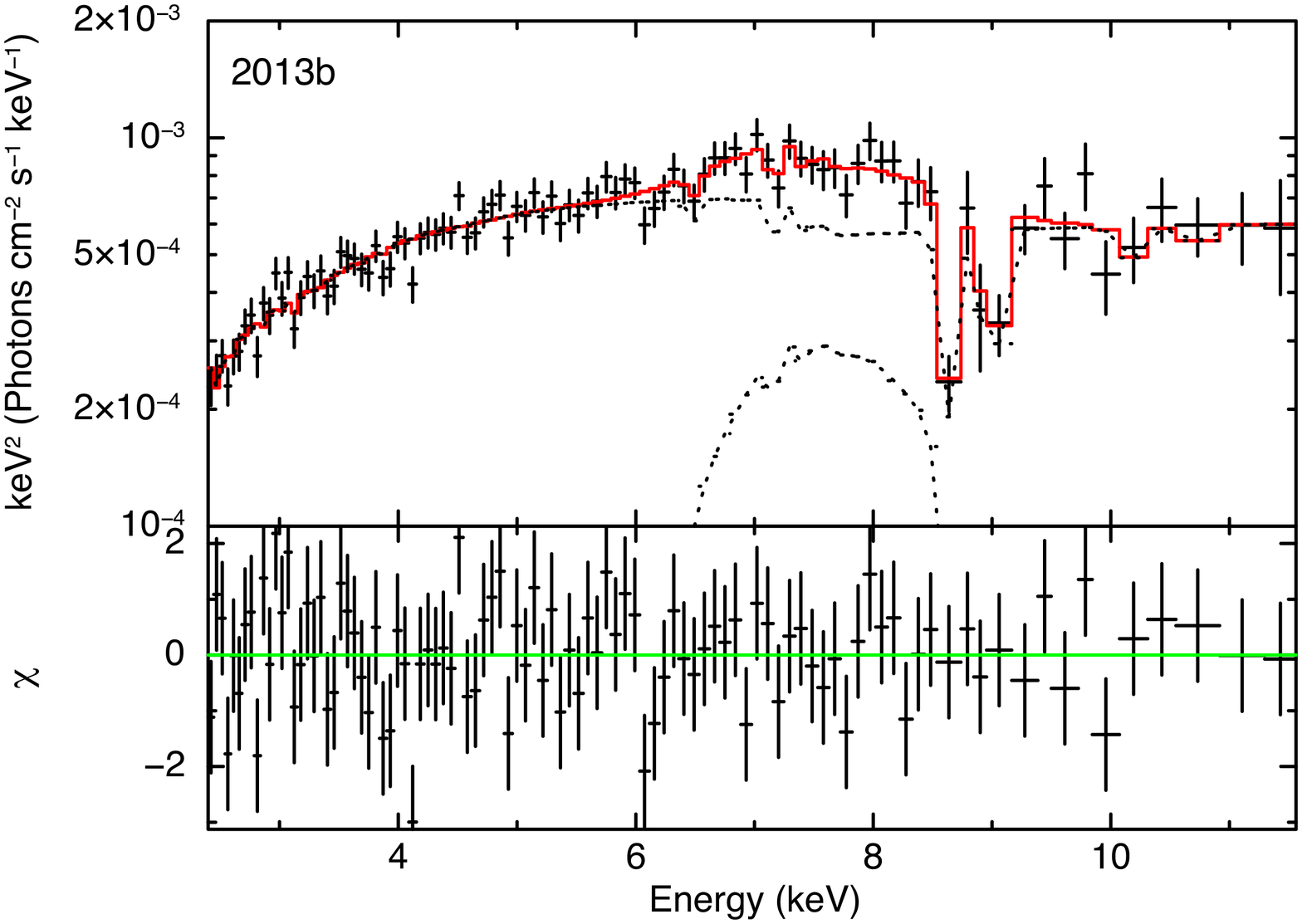}
\includegraphics[width=8cm]{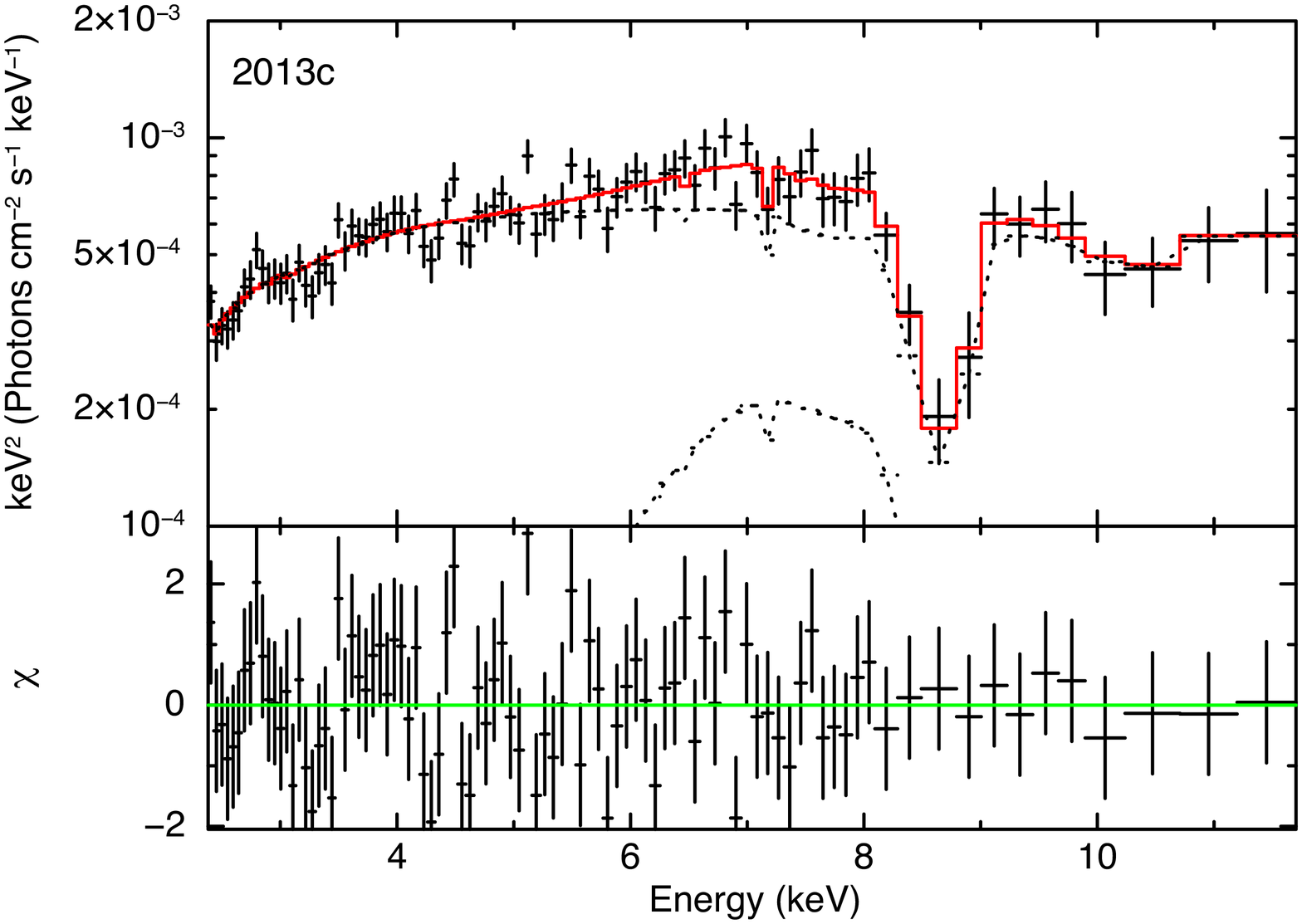}
\caption{{\it Suzaku} spectra fitted with {\sc kabs*zxipcf*powerlaw}.}
\label{suzakuspec}
\end{figure*}

Here, we assume that the intrinsic spectral shape is same as the 2007
observation, and only additional absorption makes spectral
difference. Hence we model the continuum spectra by a powerlaw model
with photon index $\Gamma=2.35$ and an ionised partial covering
absorber {\sc zxipcf}. Additional Fe absorption lines are modeled with
{\sc kabs} models.  The best fit parameters are listed in Table
\ref{paramall} and the spectra are shown in Fig. \ref{suzakuspec}.  

While the absorption lines are indeed strongest in one of the spectra
with the strongest low energy absorption (2013c, cyan in
Fig. \ref{suzakuspec}) there is not a one-to-one correlation. The
equivalent widths of absorption lines vary by more than a factor of 2
in 2013 data, while the continuum absorption is rather similar (2013a,
b and c i.e. green blue and cyan in
Fig. \ref{suzakuspec}). Conversely, the absorption line equivalent
width in 2013a (green in Fig. \ref{suzakuspec}) is significantly less
than that in the 2007 (unabsorbed) data. Thus the continuum shape
change is not directly correlated with the wind, and is hence is
unlikely to arise from a decrease in the ionisation state of the
entire wind structure. Instead, it more probably represents an
additional absorbing cloud along the line of sight. 

This cloud could be either be between the continuum source and the
wind i.e. the wind also sees the same change in illuminating spectrum
as we do, or it could be between the wind and us, in which case the
wind sees the original, unabsorbed ionising continuum.  We use XSTAR
to see if the data can distinguish between these two absorber
locations. However, the observed H--like to He--like ratio is mainly
determined by hard X-ray illumination, and this is not dramatically
changed by the absorber. Hence the current data are not able to locate
the additional absorption, and so we assume that it is outside of the 
wind, and that the wind sees the unobscured continuum. 

We note that similar, long lived, external absorption is clearly seen
in NGC~5548 \citep{Kaastra2014}, though this is typically much lower
ionisation with $\log\xi\sim-0.5$ compared to the $\log\xi\sim 2$
required by the 2013 data. This higher ionisation is caused by
K$\alpha$ ($\sim6.4$~keV) and K$\beta$ ($\sim7.1$~keV) absorptionn
lines from moderately ionised Fe ions, which are (marginally) seen in
our data (see Fig.\ref{suzakuspec}).

\begin{table*}
\caption{Spectral parameters for all {\it Suzaku} observations}
\begin{center}
\begin{tabular}{llcccc}
\hline\hline
Model Component & Fit Parameter & \multicolumn{4}{c}{Value (90\% error)}\\
\hline
& & 2011 & 2013a & 2013b & 2013c\\
\hline
Partial covering absorber 	& ${N_H}$ ($10^{22}$~cm$^{-2}$) & $1.33^{+7.83}_{-1.03}$ & $23.27^{+5.90}_{-7.00}$ & $13.67^{+2.80}_{-3.70}$ & $10.90^{+2.47}_{-5.62}$\\
		& $\log\xi$ & $-0.74 (<2.35)$ & $2.35^{+0.16}_{-0.20}$ & $2.01^{+0.17}_{-0.70}$ & $1.92^{+0.26}_{-0.59}$\\
		& $f_{cov}$ & $0.78 (>0.33)$ & $0.87^{+0.12}_{-0.08}$ & $1.00 (>0.90)$ & $0.89 (>0.84)$\\
		
Powerlaw	& $\Gamma$ & \multicolumn{4}{c}{$2.35$ (fix)}\\
		& $F_\mathrm{2-10~keV}$ ($10^{-12}$~erg~s$^{-1}$~cm$^{-2}$) & $3.14^{+0.36}_{-0.47}$ & $3.27^{+0.43}_{-0.21}$ & $2.45^{+0.16}_{-0.19}$ & $2.21^{+0.30}_{-0.49}$\\
		& $L_\mathrm{2-10~keV}$ ($10^{44}$~erg~s$^{-1}$) & $2.93^{+0.34}_{-0.44}$ & $3.06^{+0.40}_{-0.20}$ & $2.29^{+0.15}_{-0.18}$ & $2.07^{+0.28}_{-0.46}$\\
				
FeXXV He$\alpha$	&$v_{out}$ & $0.248^{+0.007}_{-0.007}c$ & $0.224^{+0.035}_{-0.019}c$ & $0.250^{+0.009}_{-0.009}c$ & $0.223^{+0.014}_{-0.021}c$\\
		& $kT$ (keV) & 474 (fix) & 474 (fix) & $2391 (<46431)$ & $11503 (<38843)$ \\
		& Natom ($10^{18}$) & $2.33^{+7.00}_{-1.47}$ & $1.01 (<2.02)$ & $2.48^{+3.48}_{-1.31}$ & $2.08 (<6.42)$\\
		& EW (keV) & $0.088$ & $0.057$ & $0.135$ & $0.125$\\

FeXXVI Ly$\alpha$	&$v_{out}$ & \multicolumn{4}{c}{tied to FeXXV}\\
		& $kT$ (keV) & \multicolumn{4}{c}{tied to FeXXV}\\
		& Natom ($10^{18}$) & $6.19^{+24.14}_{-4.25}$ & $0.59 (<3.72)$ & $4.00^{+21.09}_{-2.96}$ & $11.98 (<20.41)$\\
		& EW (keV) & $0.101$ & $0.025$ & $0.126$ & $0.232$\\

Emission	& LineE (keV) & $6.35^{+0.57}_{-1.29}$ & $7.54^{+0.24}_{-0.22}$ & $7.49^{+0.20}_{-0.18}$ & $7.08^{+0.47}_{-0.82}$\\
		& $\sigma$ (keV) & $1.43^{+1.15}_{-0.53}$ & $0.87^{+0.27}_{-0.23}$ & $0.82^{+0.29}_{-0.20}$ & $1.19^{+1.00}_{-0.52}$\\
		& EW (keV) & $0.605 (<1.152)$ & $0.667^{+0.423}_{-0.218}$ & $0.762^{+0.320}_{-0.234}$ & $0.734^{+0.757}_{-0.709}$\\

Fit statistics	     & $\chi^2$/dof	& 82.85/90	& 80.72/95	& 87.35/88	& 85.40/82\\
			     & Null probability	& 0.69		& 0.85		& 0.50		& 0.38\\
		     	     & $\chi^2$/dof for 6.5--10.0~keV & 19.81/13 & 12.84/18 & 13.02/11 & 2.26/5\\
\hline
\end{tabular}
\end{center}
\label{paramall}
\end{table*}%

\subsection{MONACO simulations}
\begin{figure*}
\centering
\includegraphics[width=8cm]{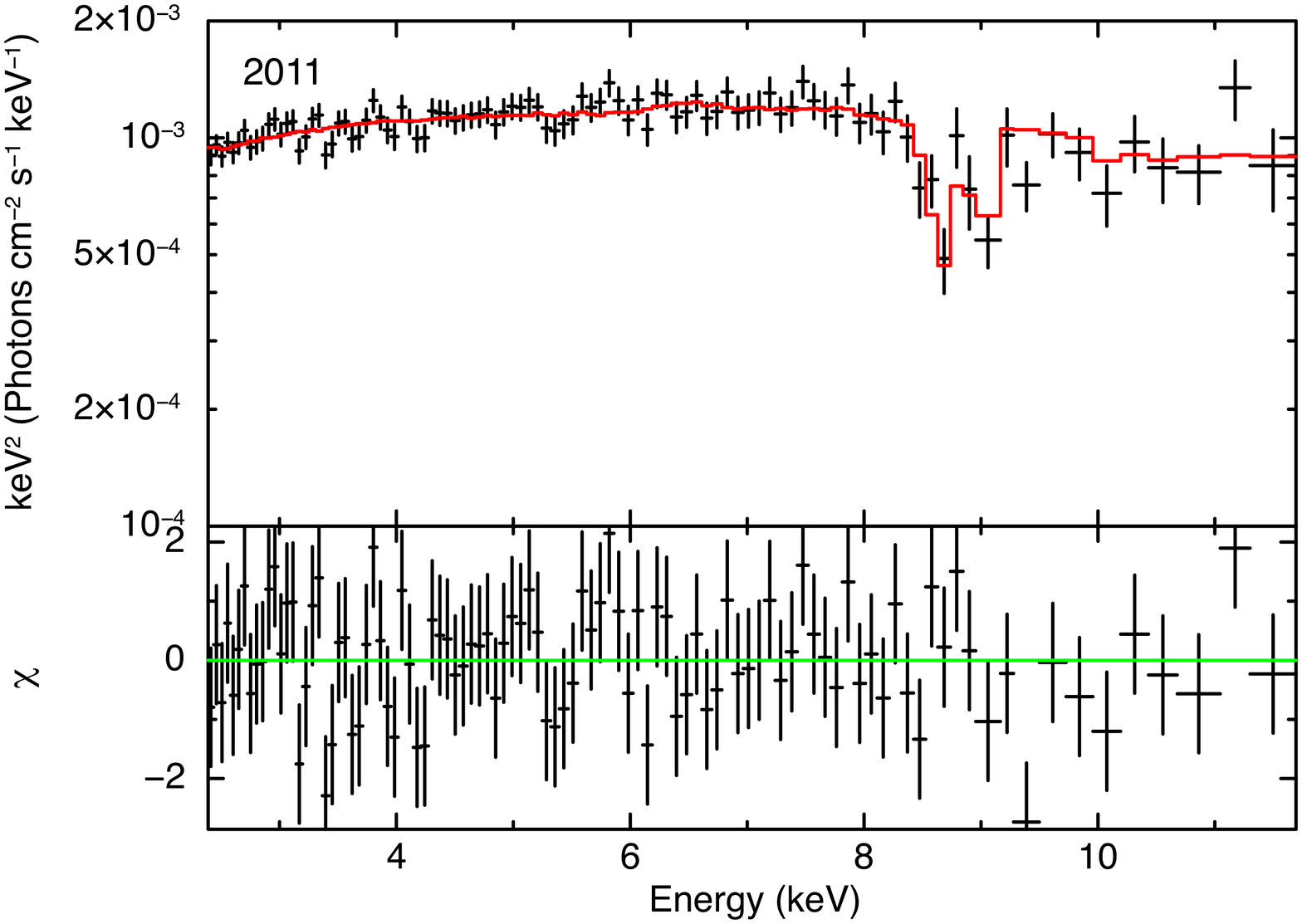}
\includegraphics[width=8cm]{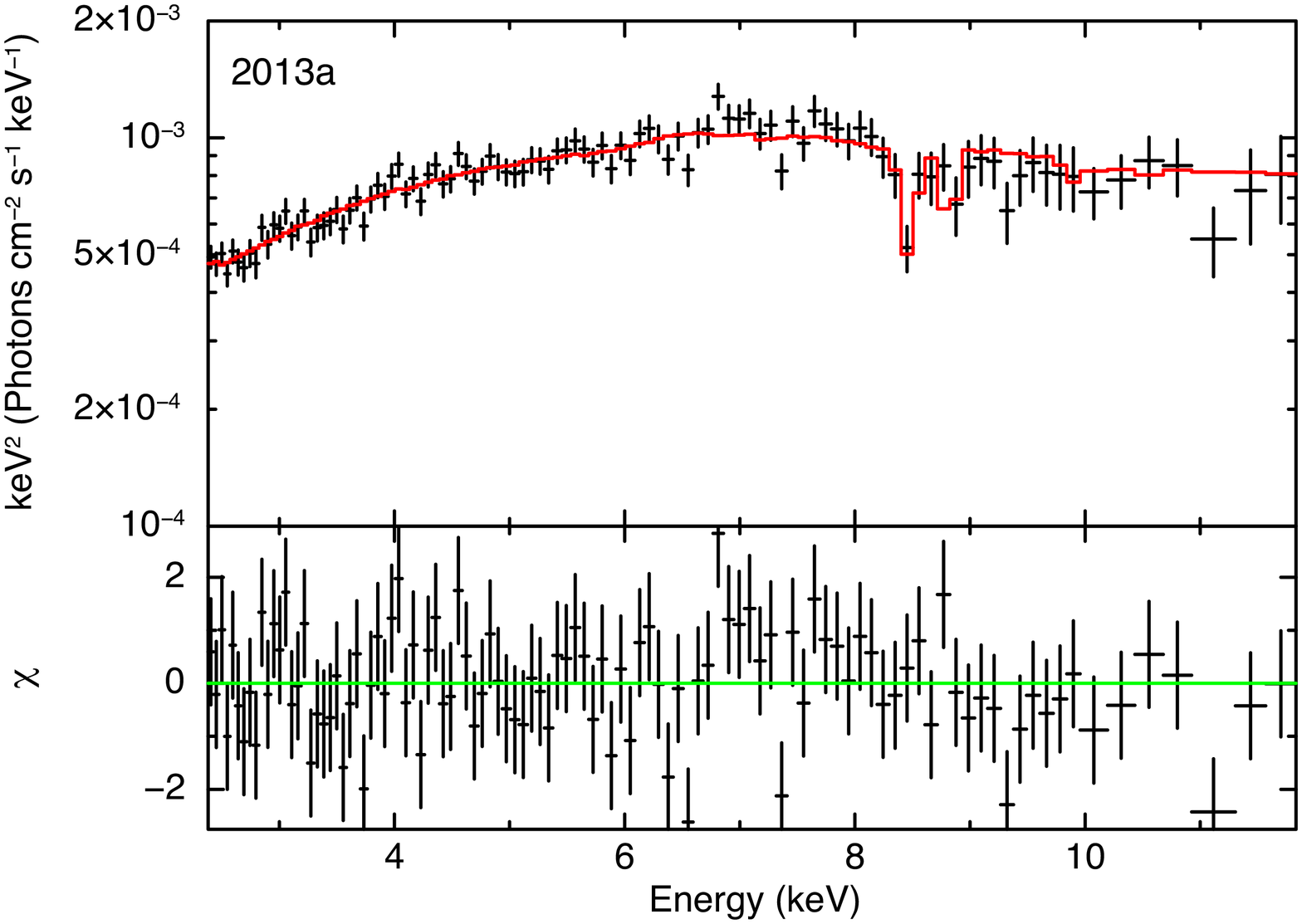}
\includegraphics[width=8cm]{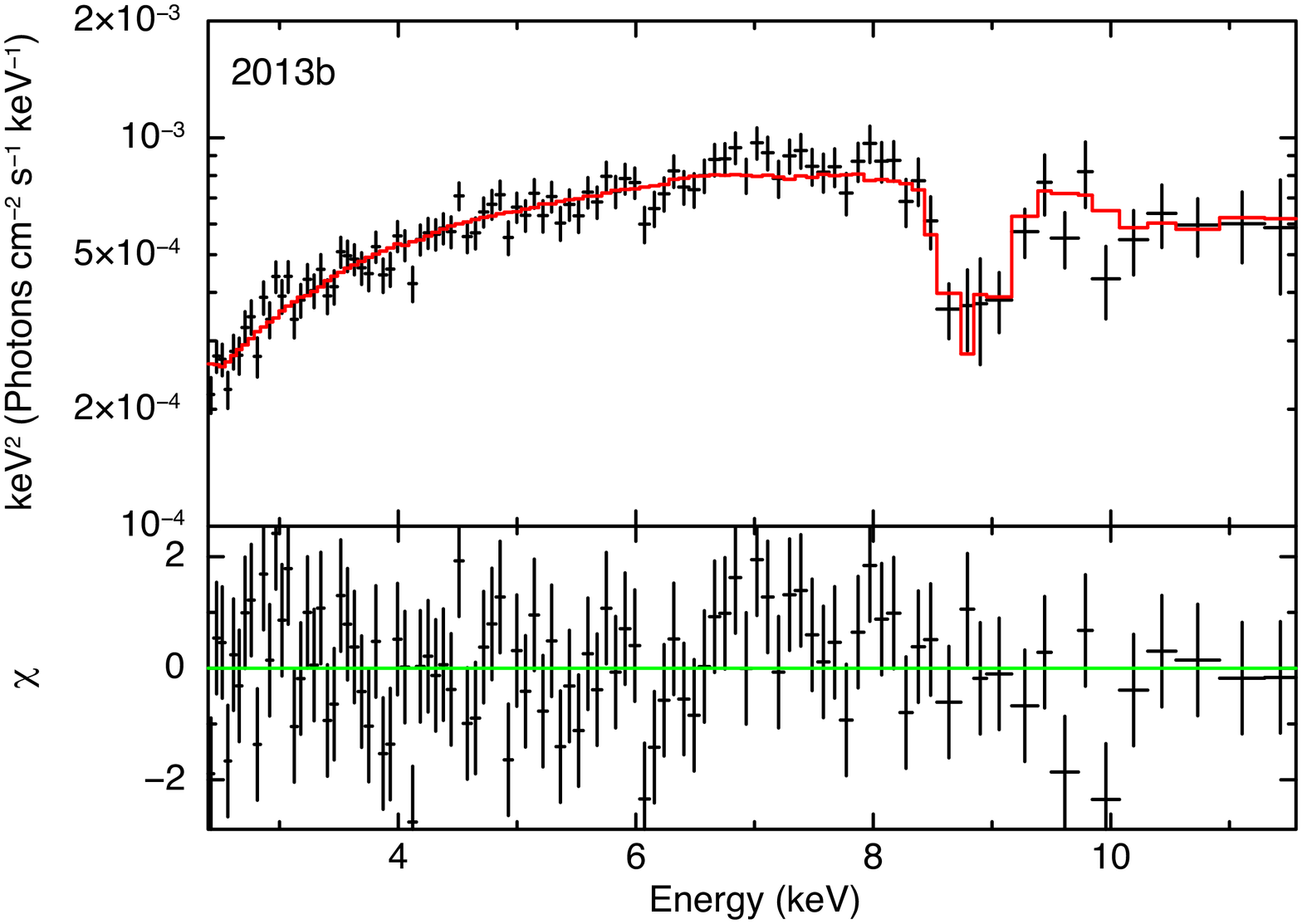}
\includegraphics[width=8cm]{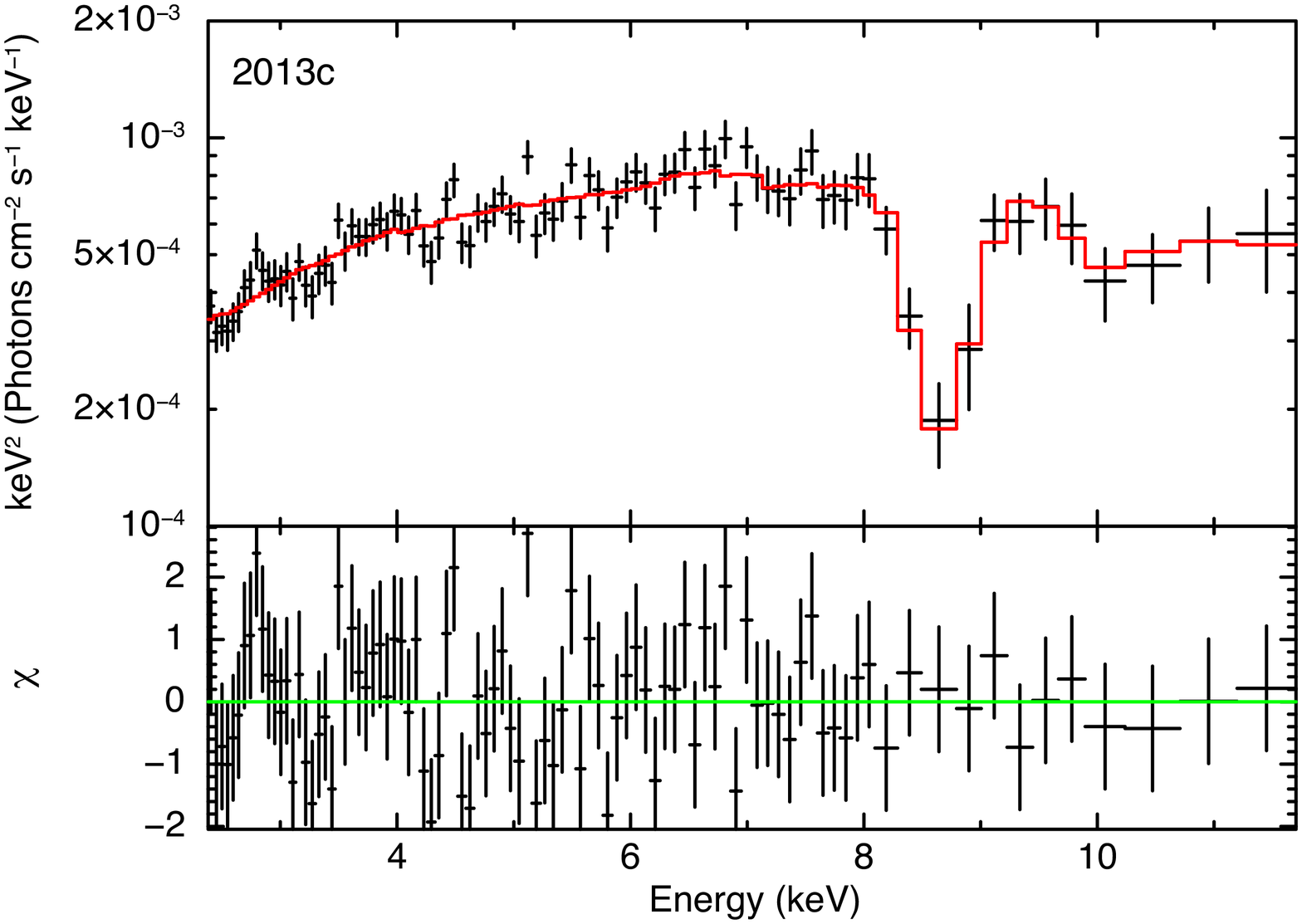}
\caption{MONACO model and {\it Suzaku} spectra for 2011, 2013a, 2013b and 2013c. All spectra are shown in the rest frame of PDS~456. }
\label{monacospec}
\end{figure*}

\begin{table*}
\caption{MONACO parameters for all {\it Suzaku} observations}
\begin{center}
\begin{tabular}{llccccc}
\hline\hline
 & Parameter & \multicolumn{5}{c}{Value}\\
\hline
& & 2007& 2011 & 2013a & 2013b & 2013c\\
\hline
MONACO wind		 & $\dot{M}_{wind}$ ($\mathrm{M_\odot}$~yr$^{-1}$) & 10 & 7 & 8 & 7 & 9\\
				 & $v_{out}$ & $0.308c$$^a$ & $0.263c$ & $0.237c$ & $0.274c$ & $0.259c$\\
				 & $\theta_{min}$ & $45^\circ$ & $43.6^\circ$ & $45.7^\circ$ & $41.6^\circ$ & $40.6^\circ$\\
				 & $\theta_{incl}$ & ${47.3^{+0.7}_{-0.6}}^\circ$ & ${48.6^{+1.1}_{-0.9}}^\circ$ & ${47.3^{+0.8}_{-1.6}}^\circ$ & ${48.0^{+1.3}_{-1.2}}^\circ$ & ${48.4^{+1.6}_{-1.4}}^\circ$\\
				 
Continuum spectra	& $N_H$ ($10^{22}$~cm$^{-2}$) & --- & $4.1^{+11.1}_{-1.1}$ & $6.0^{+5.5}_{-1.8}$ & $5.0^{+1.4}_{-1.1}$ & $9.9^{+6.4}_{-5.4}$\\
				& $\log\xi$ & --- & $-0.57 (<2.28)$ & $-0.39 (<0.31)$ & $-0.85 (<-0.26)$ & $0.27 (<1.82)$\\
				& $f_{cov}$ & --- & $0.56^{+0.07}_{-0.12}$ & $0.78^{+0.06}_{-0.09}$ & $0.91^{+0.07}_{-0.08}$ & $0.75^{0.13}_{-0.03}$\\
				& $\Gamma$ & $2.33^{+0.02}_{-0.02}$ & \multicolumn{4}{c}{$2.35$ (fix)}\\
Fit statistics	     & $\chi^2$/dof	& 106.49/105 & 89.20/95	& 102.40/100	& 105.41/94	& 89.07/88\\
			     & Null probability & 0.44	& 0.65		& 0.41		& 0.20		& 0.45\\
		     	     & $\chi^2$/dof for 6.5--10.0~keV & 21.86/27 & 24.08/18 & 21.22/23 & 19.33/17 & 3.44/11\\
\hline
\end{tabular}
\end{center}
\begin{flushleft}
$^a$ We simulate with $v_{out}=0.3c$, and then shift the spectrum.
\end{flushleft}
\label{monacoparam}
\end{table*}%

In order to determine the simulation parameters, we compared the 6.5--10.0 keV spectra of the observations between 2011 and 2013 with the model with $L=1\times10^{44}$~erg~s$^{-1}$, $v_\infty=0.3c$, $v_0=v_{turb}=1000$~km~s$^{-1}$, $\beta=1$ and $R_{min}=20R_g$.
We optimize 3 parameters of mass outflow rate $\dot{M}_{wind}$, inclination angle $\theta_{incl}$ and wind velocity $v_{out}$. Since $v_{out}$ works like redshift $z$ for absorption lines, we use $z$ instead of $v_{out}$. We simulate 4 grids of mass outflow rates, $15$~$\mathrm{M_\odot}$~yr$^{-1}$, $10$~$\mathrm{M_\odot}$~yr$^{-1}$, $8$~$\mathrm{M_\odot}$~yr$^{-1}$ and $3$~$\mathrm{M_\odot}$~yr$^{-1}$.
The geometrical parameter $\theta_{min}$ is fixed at $45^\circ$ because it doesn't have large effect on the absorption line features.

As the results, the fit statistics are best with $\dot{M}_{wind}=8$, 10, 8, $10$~$\mathrm{M_\odot}$~yr$^{-1}$, respectively for 2011, 2013a, 2013b and 2013c observations. Although $\dot{M}_{wind}=8$, 10, $15$~$\mathrm{M_\odot}$~yr$^{-1}$ gives comparably good fit for any observations, we choose the best fit value of $\dot{M}_{wind}$. For these mass outflow rates, the best fit values of redshift are $z=0.232, 0.267, 0.218, 0.238$ respectively, which corresponds to $v=0.263\pm0.006c$, $0.237\pm0.008c$, $0.274\pm0.006c$ and $0.259\pm0.007c$.
Here, the obtained values of mass outflow rate should be corrected by the outflow velocity because the outflow velocity is assumed to be $0.3c$ in the simulations. 
According to Eq. \ref{mass0}, the mass outflow rate $\dot{M}$ is proportional to the density $n$ and the outflow velocity $v$ as $\dot{M}\propto nv$. Since the density determines the ionisation structure and the absorption column, the density in the simulation $n_{sim}$ has to equal to that in the observed spectra $n_{obs}$. 
Thus, the corrected mass outflow rate is $\dot{M}_{obs}\simeq\dot{M}_{sim}v_{obs}/v_{sim}$, and the best fit values of the mass outflow rate become $\dot{M}_{wind}\simeq7$, 8, 7, 9~$\mathrm{M_\odot}$~yr$^{-1}$ respectively. 
For the absorption lines, the change of inclination angle is interpreted as the change of opening angle of the wind and/or change of $\theta_{min}$. 
Here, it is assumed that the geometrical parameter $\theta_{min}$ equals to $45^\circ$ in the 2007 observation.

The comparison between the observed spectra and our simulation models is shown in Fig. \ref{monacospec}.
All simulation parameters are listed in Table \ref{monacoparam}.
The observed time variability of the wind
could be caused by the hydrodynamic instability of a UV line driven disk wind as seen in 
\cite{Proga2004} and \cite{Nomura2014}. Variability of the wind on even shorter timescales is 
discussed by \cite{Gofford2014}.

\section{Discussion}

Table \ref{monacoparam} shows that the best fit values of mass
outflow rate of winds in PDS~456 of
$\dot{M}_{wind}\simeq7$--$10$~$\mathrm{M_\odot}$~yr$^{-1}$, roughly
30\% of the total mass inflow rate as traced by the optical emission
from the outer disc. The kinetic energy and momentum of the wind are
close to that provided by the radiation field \citep{Gofford2013, Gofford2014}, pointing to the importance of radiative driving in launching
and accelerating the wind.  However, the mechanism for this is
unclear. UV line driving results in powerful winds from the UV bright
O stars and disc accreting white dwarfs, but the X-rays which
accompany the bright UV discs in AGN strongly suppress the wind
through overionisation \citep{Proga2004}. The UFO's are so
highly ionised that there is no UV or even soft X-ray opacity left, so
UV line driving cannot be accelerating the highly ionised material
which we see \citep{Higginbottom2014}.

However, here we suggest a solution to this issue. UV line driving
could be launching and accelerating the material from the disc. As it
rises higher it is pushed outwards and ionised by the harder UV and
X-ray radiation from the inner disc. The UV opacity in then mostly
on the vertically rising part of the wind, which is outside of our 
line of sight (see e.g. the wind geometries in \citealt{Risaliti2010}; \citealt{Nomura2013}).

We can estimate the effect of this in PDS~456. 
Without mass loss, such a disc should have
$L(20-30R_g)=0.64 L(6-20R_g)$, so reducing the inner disc luminosity
by 2/3 to account for the smaller mass accretion rate gives
$L(20-30R_g)\approx L(6-20R_g)$.  Assuming that the wind is launched
vertically by the disc luminosity from $20-30R_g$, and pushed sideways
by the inner disc luminosity from $6-20R_g$ gives an estimate for
$\theta_0\sim 45^\circ$, the angle the wind makes to the disc
normal. This is even more convincingly close to our fiducial geometry
than with the standard (no mass loss in a wind) disc (see Section \ref{sec:param}).

\begin{table}
\caption{Full numerical calculation of UV-line driven winds. All accretion rates are in units of the Eddington accretion rate.}
\begin{center}
\begin{tabular}{lcccc}
\hline\hline
$L/L_{Edd}$ & $a$$^a$ & $\dot{M}_{in}$$^b$ & $\dot{M}$$^c$ & $\dot{M}_{wind}$$^d$\\
\hline
\multirow{2}{*}{0.3} & 0   &    0.302   &   0.515  &   0.21\\
& 0.9  &   0.302   &   1.174  &   0.87\\
\hline
\multirow{2}{*}{1.0}  & 0  &     1.007   &   2.282  &   1.09\\
 & 0.9  &   1.007  &    5.588   &  4.58\\
\hline
\end{tabular}
\end{center}
\begin{flushleft}
$^a$ Black hole spin parameter\\
$^b$ The amount of mass that is actually accreted\\
$^c$ The accretion rate at large radius before the outflow set in\\
$^d$ Mass outflow rate\\
\end{flushleft}
\label{UVcalc}
\end{table}%

\cite{Laor2014} have done a much more exact calculation of the
effect of mass loss on the disc structure. Their models include the
energy to power the wind to its local escape velocity ($\epsilon=1$)
on the structrue of the remaining disc, as well as the effect of
angular momentum losses and decrease in mass accretion rate.  They
parameterize the mass loss rate from each surface element of the disc
by using observed O star winds i.e. they assume that the winds are UV
line driven, and scale for the different gravity ($g$)
conditions. This gives a surface density mass loss rate of
$\dot{\Sigma}\propto F^{2.32}/g^{1.11}$, where $F\propto T^4$ is the
local surface flux. However, O stars only span a rather small range in
temperature, from $2.8-5\times 10^4$~K \citep{Howarth1989}, so
this relation only formally holds for this range. Nonetheless, this is
close to the disc temperatures expected for such a high mass black
hole, so the \cite{Laor2014} results should be applicable.
Tab. \ref{UVcalc} shows the full numerical calculation of UV-line
driven winds (Shane Davis, private communication). This calculation is
done for $10^{9}$~$\mathrm{M_\odot}$ black hole for $a_*=0$ and $0.9$,
accreting at $L/L_{Edd}=0.3$ and $1$. These show that mass loss rates
of 30-50\% of the mass inflow rate are expected from UV line driven
disc winds assuming that the central X-ray flux does not overionise
the wind.

The X-ray power then becomes critically important, and AGN are
observed to show an anti-correlation of X-ray flux with $L/L_{Edd}$
(\citealt{Vasudevan2007}; \citealt{Jin2012a, Jin2012b}; \citealt{Done2012} see
their Fig 8a and b). While the underlying reason for this is not well
understood, it is clear that as a source approaches $L_{Edd}$ then
radiation pressure alone means that winds become important, while the
drop in X-ray luminosity means that UV line driving becomes more
probable since the X-ray ionisation drops.  This combination of
continuum and UV line driving seems the most likely way to drive the
most powerful winds.

\begin{figure}
\centering
\includegraphics[angle=-90,width=8cm]{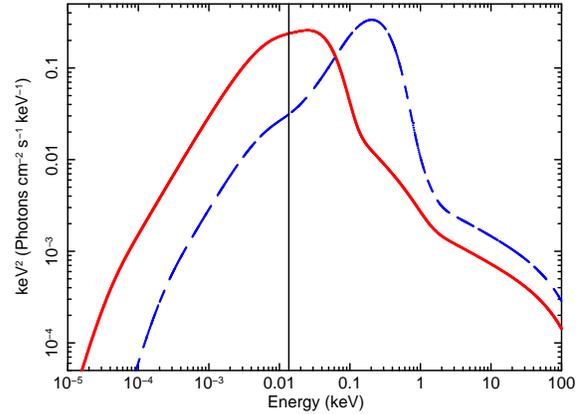}
\caption{Predicted spectral energy distributions for spin zero black holes of mass $10^6$ (blue dashed) and $10^9\mathrm{M_\odot}$ (red). The black vertical line is at 13.6~eV Hydrogen
ionisation point which corresponds to the UV line driving bandpass. Here, we use {\sc optxagnf} \citep{Done2012} and assume $L/L_{Edd}=1$. }
\label{bhmass}
\end{figure}

This predicts that fast winds should be suppressed in lower
$L/L_{Edd}$ objects, as $L/L_{Edd}\ll 1$ means that the wind cannot be
powered by continuum driving (definition of the Eddington limit) and
the higher X-ray flux means that UV line driving is strongly
suppressed. It also predicts that the fastest winds should be seen in
the highest mass objects with $L/L_{Edd}\sim 1$ as these are the ones
where the disc luminosity peaks in the UV rather than the far UV/soft
X-rays, where the disc itself contributes to overionising the wind.
Fig \ref{bhmass} shows the predicted spectral energy distributions for
$L/L_{Edd}=1$ for Schwarzchild black holes of mass $10^6$ (blue) and
$10^9\mathrm{M_\odot}$ (red). These assume that the accretion energy
is dissipated in a standard (constant mass inflow rate) disc, and
thermalises to a (colour temperature corrected) blackbody down to
$10R_g$, and that 30\% of the accretion energy below this powers a
tail to high energies with $\Gamma=2.4$, while the remainder powers a
low temperature, optically thick corona ($kT_e=0.2$, $\tau=15$: see
\citealt{Done2012}). The black vertical line marks the 13.6~eV Hydrogen
ionisation point. A blackbody at O star temperatures will peak in the
10-18~eV range, so this indicates the UV line driving bandpass.
Clearly the disc for the more massive black hole will have much
stronger UV line driving that the less massive one.  Simply assigning
all of the disc luminosity to a UV band as is often done in
hydrodynamic calculations to make them numerically tractable \citep{Proga2004,Nomura2014} does not include this mass
dependence, so may overestimate the wind mass loss rates for lower
mass AGN (e.g. \citealt{Laor2014}).

Thus we expect the most powerful winds to be powered by a combination
of continuum and UV line driving, and for these winds to be found in
the most massive AGN. This is clearly the case, with the winds in
PDS~456 and APM~08279+5255, both high mass ($>10^9\mathrm{M_\odot}$) black
holes at $L\sim L_{Edd}$, standing out as by far the highest velocity,
highest mass loss rate objects \citep{Tombesi2010,Gofford2013}. 
We will fit the wind in APM~08279+5255 in a subsequent paper.

\section{Conclusions}

We show that the geometry and energetics of wind in PDS~456 can be
constrained using our new combined Monte-Carlo and ionisation code,
MONACO. The code treats only H and He-like ions, but this makes it
fast enough that we can explore parameter space for highly ionised
winds, where this approximation is appropriate.

Our simulations successfully reproduce all the {\it Suzaku} observations of PDS~456. In particular, we can explain the time variability of the wind spectra within several weeks observed in 2013 in our modelling framework.
Most of the fundamental parameters are kept constant in our simulations but wind velocity and relative angle between the line of sight and wind direction are sightly changed.

From our simulations, we find that the best fit values of mass outflow rate of winds in PDS~456 are 7--10~$\mathrm{M_\odot}$~yr$^{-1}$, corresponding to $\sim30$\% of the total mass inflow rate.
According to full numerical calculation of UV-line driven winds done by Laor \& Davis, these results can match the properties of UV line driven disc wind models.
The wind is vertically accelerated by the UV emission from the disc before it is pushed sideways by the inner disc emission and ionised by the central X-ray source. This mechanism works most efficiently in high mass AGN, as their discs peak in the UV.
Observations also show that as AGN approach Eddington, the fraction of X-ray luminosity decreases. This helps the wind not to be overionised, as well as giving extra acceleration to the wind from continuum radiation driving. Thus the most extreme outflows are predicted to be observed in high mass, high Eddington fraction AGN.

\section*{ACKNOWLEDGMENTS}
C.D. thanks Shane Davis for the calculations of the UV line driven
disk winds shown in section 6, and for multiple useful conversations
about disks and winds.  K.H. is supported by the Japan Society for the
Promotion of Science (JSPS) Research Fellowship for Young
Scientists. We thank the referee for their comments which improved the
structure of the paper.

\bibliographystyle{mn2e}
\bibliography{ref} 

\appendix
\section{Parameter dependence}

\begin{figure*}
\centering
\includegraphics[width=16cm]{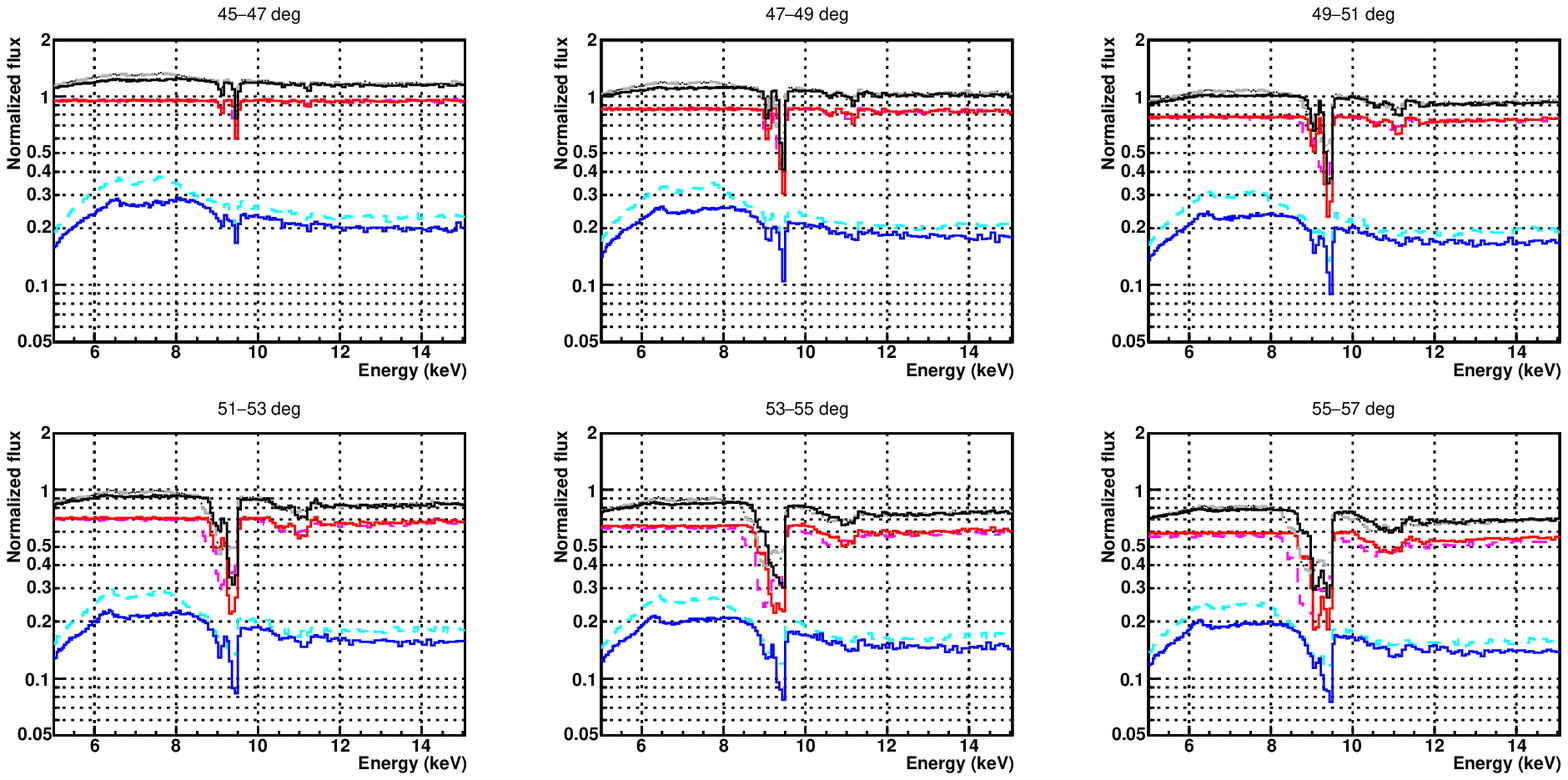}
\caption{Dependence on the velocity law. The grey, magenta and cyan
lines show the fiducial parameter simulation with 
$L=4\times10^{44}$~erg~s$^{-1}$, $\dot{M}=10$~$\mathrm{M_\odot}$~yr$^{-1}$, while the 
black, red and blue dashed curves show the same parameters except with an
acceleration law $\beta=0.5$.
}
\label{monaco_l4_m10_vt_b05}
\end{figure*}

\begin{figure*}
\centering
\includegraphics[width=16cm]{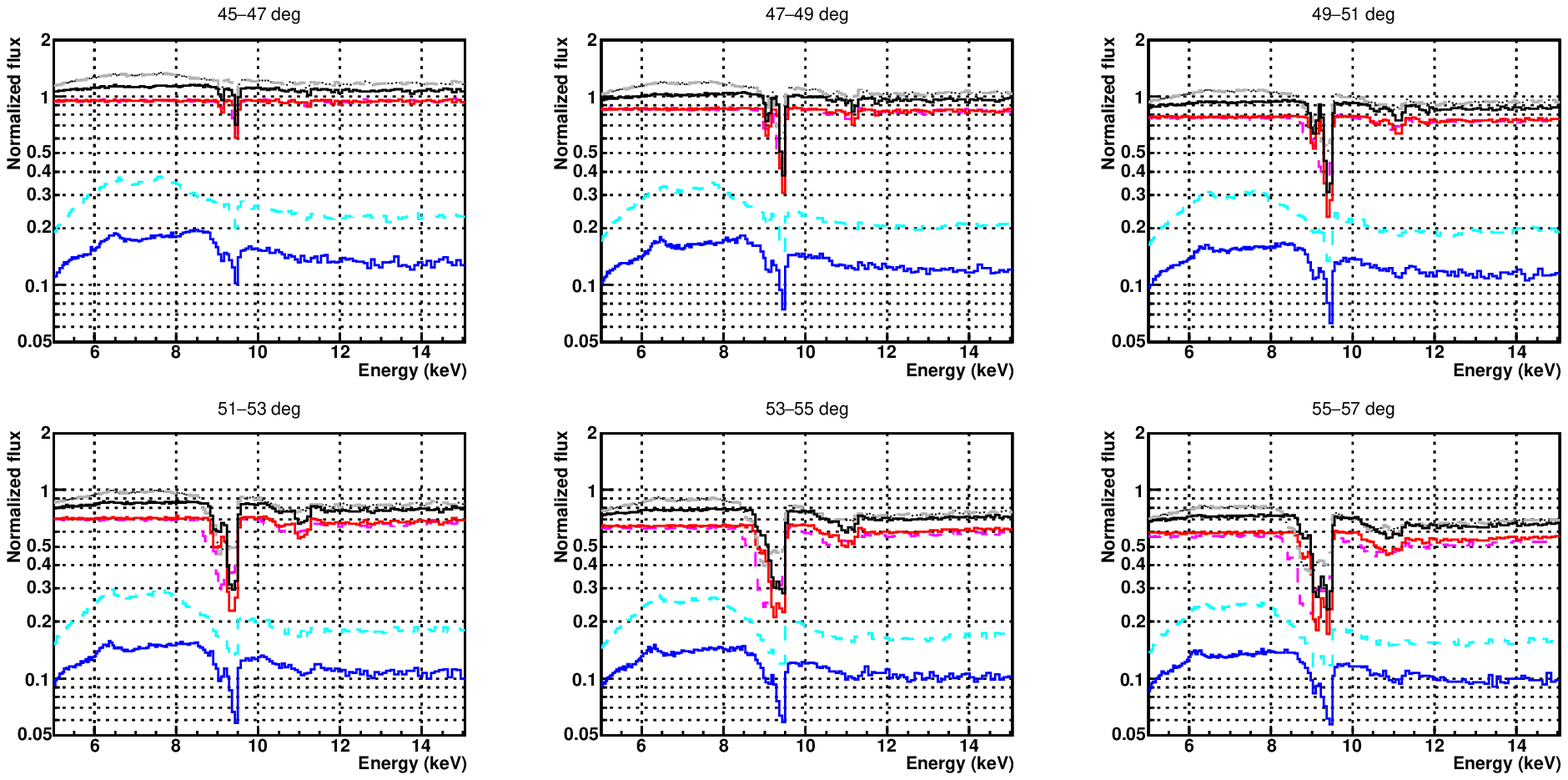}
\caption{Dependence on the initial velocity. The grey, magenta and cyan
lines show the fiducial parameter simulation with 
$L=4\times10^{44}$~erg~s$^{-1}$, $\dot{M}=10$~$\mathrm{M_\odot}$~yr$^{-1}$, while the 
black, red and blue dashed curves show the same parameters except with an initial velocity $v=0.15c$.
}
\label{monaco_l4_m10_v015}
\end{figure*}

\begin{figure*}
\centering
\includegraphics[width=16cm]{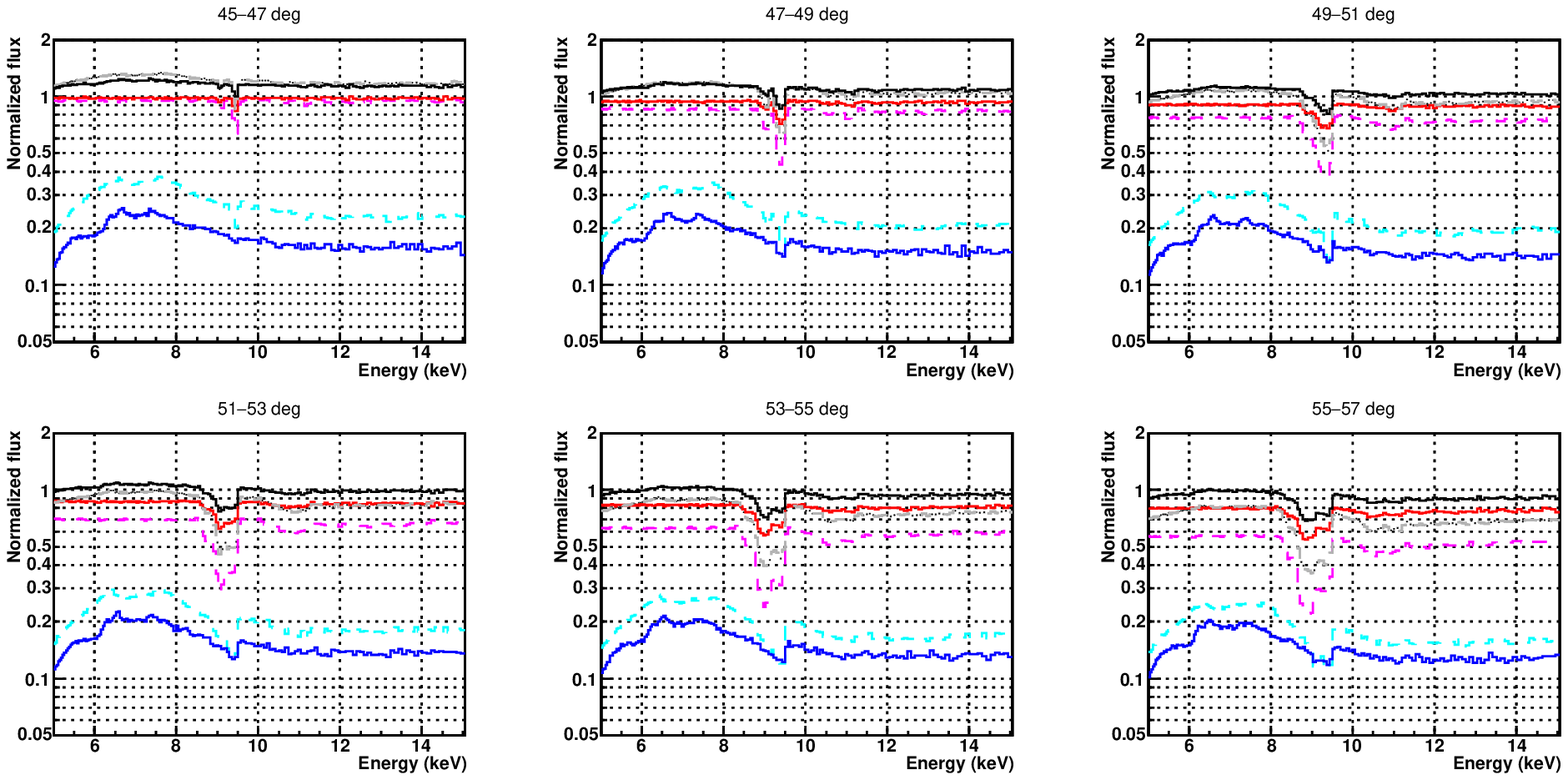}
\caption{Dependence on the launch radius. The grey, magenta and cyan
lines show the fiducial parameter simulation with 
$L=4\times10^{44}$~erg~s$^{-1}$, $\dot{M}=10$~$\mathrm{M_\odot}$~yr$^{-1}$, while the 
black, red and blue dashed curves show the same parameters except with an launch radius $50-75R_g$.
}
\label{monaco_r50_l4_m10_vt}
\end{figure*}

\begin{figure*}
\centering
\includegraphics[width=16cm]{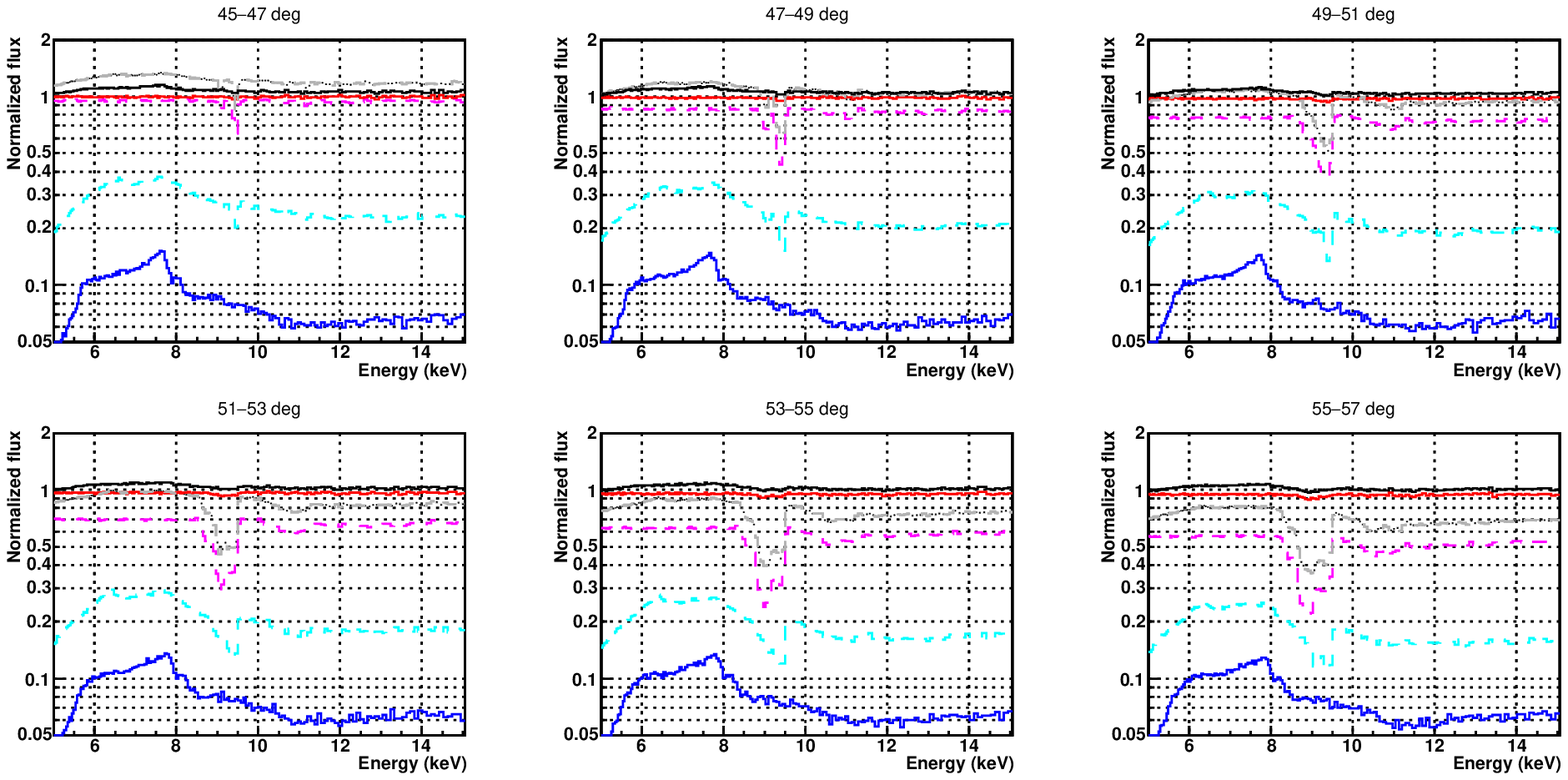}
\caption{Dependence on the mass outflow rate. The grey, magenta and cyan
lines show the fiducial parameter simulation with 
$L=4\times10^{44}$~erg~s$^{-1}$, $\dot{M}=10$~$\mathrm{M_\odot}$~yr$^{-1}$, while the 
black, red and blue dashed curves show the same parameters except with an ionising luminosity $L=1\times10^{44}$~erg~s$^{-1}$ and $\dot{M}=1$~$\mathrm{M_\odot}$~yr$^{-1}$.
}
\label{monaco_l1_m1_vt}
\end{figure*}

\begin{figure*}
\centering
\includegraphics[width=16cm]{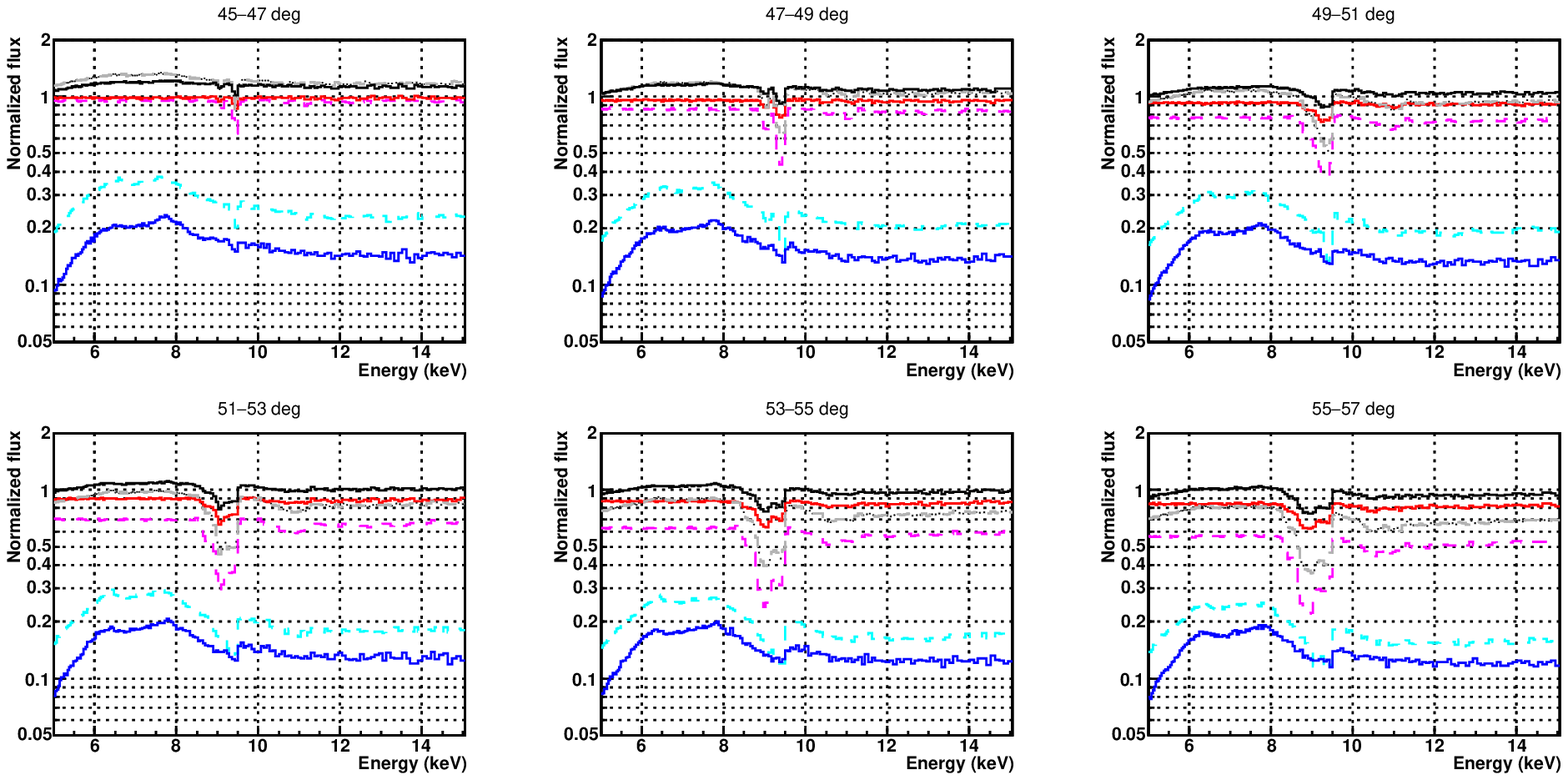}
\caption{Dependence on the mass outflow rate. The grey, magenta and cyan
lines show the fiducial parameter simulation with 
$L=4\times10^{44}$~erg~s$^{-1}$, $\dot{M}=10$~$\mathrm{M_\odot}$~yr$^{-1}$, while the 
black, red and blue dashed curves show the same parameters except with an ionising luminosity $L=1\times10^{44}$~erg~s$^{-1}$ and $\dot{M}=3$~$\mathrm{M_\odot}$~yr$^{-1}$.
}
\label{monaco_l1_m3_vt}
\end{figure*}

\begin{figure*}
\centering
\includegraphics[width=16cm]{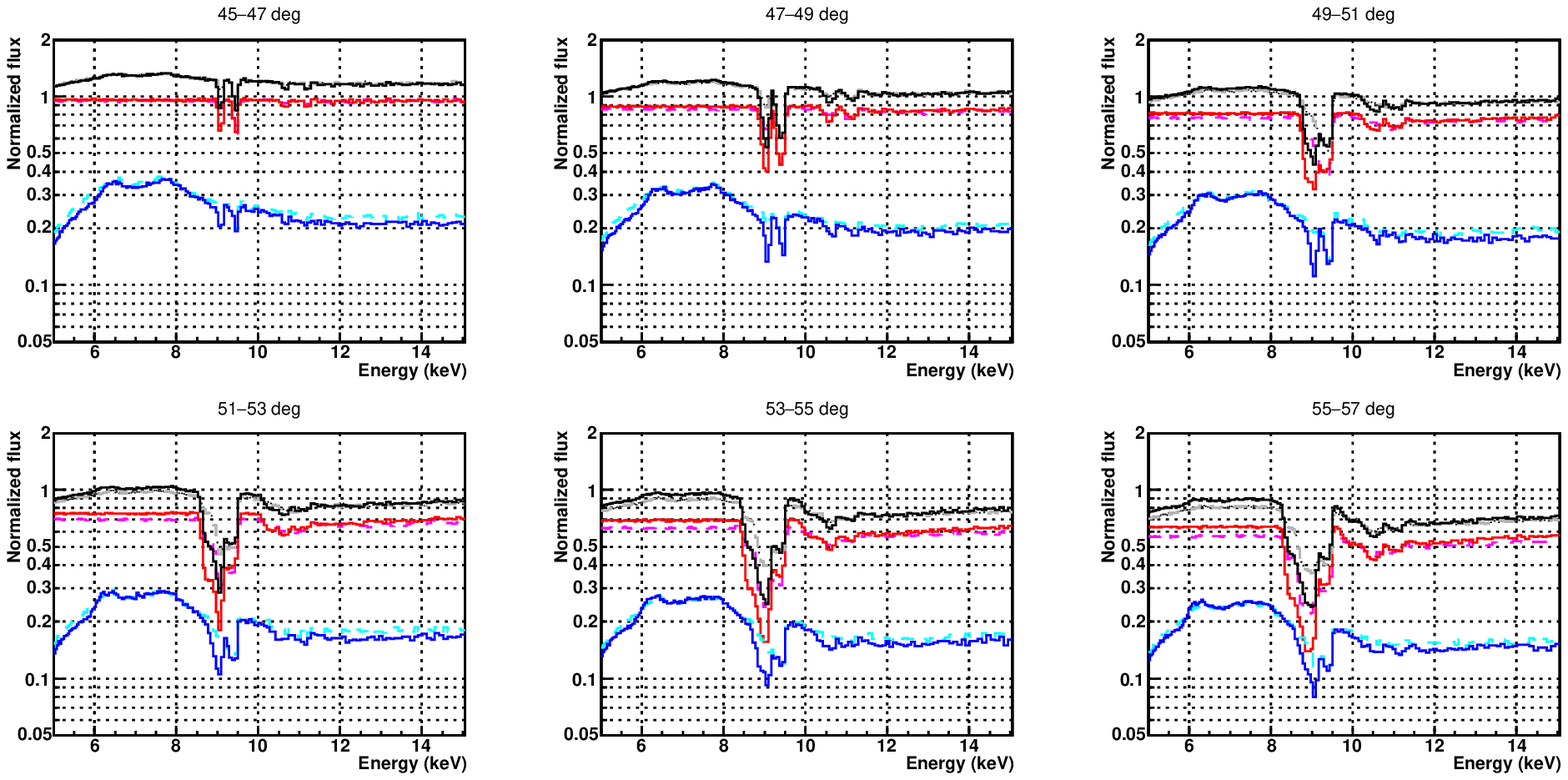}
\caption{Dependence on the mass outflow rate. The grey, magenta and cyan
lines show the fiducial parameter simulation with 
$L=4\times10^{44}$~erg~s$^{-1}$, $\dot{M}=10$~$\mathrm{M_\odot}$~yr$^{-1}$, while the 
black, red and blue dashed curves show the same parameters except with an ionising luminosity $L=1\times10^{44}$~erg~s$^{-1}$ and $\dot{M}=8$~$\mathrm{M_\odot}$~yr$^{-1}$.
}
\label{monaco_l1_m8_vt}
\end{figure*}

\end{document}